\documentclass[useAMS,usenatbib]{mn2e}
\usepackage{graphicx}
\usepackage{times}
\usepackage{natbib}
\usepackage{setspace}
\newif\ifAMStwofonts
\AMStwofontstrue

\newcommand{\be}{\begin{equation}}
\newcommand{\ee}{\end{equation}}
\newcommand{\ba}{\begin{eqnarray}}
\newcommand{\ea}{\end{eqnarray}}
\newcommand{\brr}{\begin{array}}
\newcommand{\err}{\end{array}}
\newcommand{\bc}{\begin{center}}
\newcommand{\ec}{\end{center}}

\newcommand{\kms}{km s$^{-1}$}

\newcommand{\surf}{M$_\odot$ pc$^{-2}$}
\newcommand{\ssfr}{M$_\odot$ yr$^{-1}$ kpc$^{-2}$}

\newcommand{\muppi}{{\sc muppi}}
\newcommand{\gtre}{{\sc gadget}3}

\newcommand{\mincir}{\raise
  -2.truept\hbox{\rlap{\hbox{$\sim$}}\raise5.truept \hbox{$<$}\ }}
\newcommand{\magcir}{\raise
  -2.truept\hbox{\rlap{\hbox{$\sim$}}\raise5.truept \hbox{$>$}\ }}
\newcommand{\siml}{\raise
  -2.truept\hbox{\rlap{\hbox{$\sim$}}\raise5.truept \hbox{$<$}\ }}
\newcommand{\simg}{\raise
  -2.truept\hbox{\rlap{\hbox{$\sim$}}\raise5.truept \hbox{$>$}\ }}

\title[SK relations in SPH simulations] 
{Schmidt-Kennicutt relations in
SPH simulations of disc galaxies with effective thermal feedback from SNe}

\author[P. Monaco et al.] {
Pierluigi Monaco$^{1,2}$, Giuseppe Murante$^{3,1}$, 
Stefano Borgani$^{1, 2, 4}$, Klaus Dolag$^{5,6}$\\
$^1$ Dipartimento di Fisica - Sezione di Astronomia, Universit\`a di Trieste, via Tiepolo 11, I- 34131 Trieste -- Italy (monaco, borgani@oats.inaf.it)\\ 
$^2$ INAF, Osservatorio Astronomico di Trieste, Via Tiepolo 11, I-34131 Trieste -- Italy \\
$^3$ INAF, Osservatorio Astronomico di Torino, Strada Osservatorio 20, I-10025 Pino Torinese -- Italy (murante@oato.inaf.it)\\ 
$^4$ INFN, Istituto Nazionale di Fisica Nuclare, Trieste -- Italy \\ 
$^5$ University Observatory Munich, Scheinerstr. 1, D-81679 M\"unchen -- Germany (kdolag@mpa-garching.mpg.de) \\
$^6$ Max-Planck-Institut f\"ur Astrophysik, Karl-Schwarzschild-Strasse 1, D-85748 Garching bei M\"uunchen -- Germany 
}

\begin{document}

\maketitle

\label{firstpage}

\begin{abstract}

We study several versions of the Schmidt-Kennicutt (SK) relation
obtained for isolated spiral galaxies in TreeSPH simulations run with
the {\gtre} code including the novel {\it MUlti-Phase Particle
  Integrator} (\muppi) algorithm for star formation and stellar
feedback.  This is based on a sub-resolution multi-phase treatment of
gas particles, where star formation is explicitly related to molecular
gas, and the fraction of gas in the molecular phase is computed from
hydrodynamical pressure, following a phenomenological correlation.  No
chemical evolution is included in this version of the code.  The
standard SK relation between surface densities of cold
(neutral+molecular) gas and star formation rate of simulated galaxies
shows a steepening at low gas surface densities, starting from a knee
whose position depends on disc gas fraction: for more gas-rich discs
the steepening takes place at higher surface densities.  Because gas
fraction and metallicity are typically related, this environmental
dependence  mimics the predictions of models where the formation
of $H_2$ is modulated by metallicity.  The cold gas surface density at
which HI and molecular gas surface densities equate can range from
$\sim10$ up to $34$ {\surf}.  As expected, the SK relation obtained
using molecular gas shows much smaller variations among simulations.
We find that disc pressure is not well represented by the classical
external pressure of a disc in vertical hydrostatic
equilibrium. Instead is well fit by the expression
$P_{\rm fit} = \Sigma_{\rm cold} \sigma_{\rm cold} \kappa / 6$,
where the three quantities on the right-hand side are cold gas surface
density, vertical velocity dispersion and epicyclic frequency.  When
the ``dynamical'' SK relation, i.e. the relation that uses gas surface
density divided by orbital time, is considered, we find that all of
our simulations stay on the same relation.  We interpret this as a
manifestation of the equilibrium between energy injection and
dissipation in stationary galaxy discs, when energetic feedback is
effective and pressure is represented by the expression given above.
These findings further support the idea that a realistic model of the
structure of galaxy discs should take into account energy injection by
SNe.
\end{abstract}

\begin{keywords}
galaxies: formation -- galaxies: ISM -- galaxies: kinematics and dynamics
\end{keywords}

\section{Introduction}
\label{section:intro}

In the last decade a significant step forward in the phenomenological
understanding of star formation in galaxies has been achieved, thanks
to many observational campaigns of nearby \citep[see,
  e.g.,][]{Boissier07,Kennicutt07,Walter08} and distant
\citep[e.g.][]{Bouche07,Daddi10a,Genzel10} galaxies.  In particular,
the relation between surface densities of gas and star formation rate
(hereafter SFR), the so-called Schmidt-Kennicutt relation
\citep[][hereafter standard SK]{Schmidt59,Kennicutt98}, has acquired a
higher degree of complexity, passing from a simple power law to a much
more structured and environment-dependent relation.  Here by
environment we mean the local properties, averaged over $\sim$kpc
scale, of a galaxy patch that hosts star-forming molecular clouds.  We
will call {\it standard}, {\it molecular}, {\it HI} and {\it
  dynamical} SK the relations obtained by putting on the y-axis the
surface density of SFR $\Sigma_{\rm sfr}$, and on the x-axis,
respectively, the total cold gas surface density $\Sigma_{\rm cold}$,
the molecular gas surface density $\Sigma_{\rm mol}$, the HI gas
surface density $\Sigma_{HI}$ or the total cold gas surface density
divided by the orbital time-scale, $\Sigma_{\rm cold}/\tau_{\rm orb}$,
where $\tau_{\rm orb}=2\pi r/V(r)$ ($V(r)$ being the galaxy rotation
curve). 

Wide consensus has recently been reached on the idea that the standard
SK is a reflection of the more ``fundamental'' molecular SK
\citep{Wong02,Blitz04,Blitz06}, while the HI SK is weak if not absent
\citep{Bigiel08,Bigiel10}.  Indeed, in normal (and non-barred) spiral
galaxies the fraction of gas in molecular clouds increases toward the
galaxy center, while the HI gas surface density, $\Sigma_{HI}$,
saturates at a value of $\sim10$ {\surf} and declines in the inner
kpc.  The old notion of a star formation threshold in disc galaxies
\citep{Martin01} has thus been revised to a steepening of the standard
SK at low surface densities \citep{Boissier07,Bigiel08}, driven by the
declining molecular fraction.

The gas surface density at which the transition from HI- to
molecular-dominated gas takes place, or equivalently the saturation
value of $\Sigma_{HI}$, has been proposed to be a function of galactic
environment (\citealt{Krumholz09b,Gnedin10,Papadopoulos10}; see also
\citealt{Schaye04}), with dwarf galaxies like the Magellanic Clouds
showing higher values (Bolatto et al. 2009; \citealt{Bolatto11}; see also references in
\citealt{Fumagalli10}).  This is in line with the high-redshift evidence
of low efficiency of star formation in Damped Lyman-alpha systems at
$z\sim2$ \citep{Wolfe06}, that are thought to be the external,
gas-rich, metal-poor regions of young disc galaxies.

The slope of the molecular SK relation is debated. For THINGS spiral
galaxies, \cite{Bigiel08} report a slope of $1.0\pm0.2$ when measured on 
a spatial grid of 750 pc bin size. An average value of $\sim$1 has been
confirmed very recently by \cite{Bigiel11}.  
A steeper slope, more consistent with the
canonical 1.4 value of \citealt{Kennicutt98} is reported by
\cite{Kennicutt07} for star-forming regions in M51a.
\cite{Liu11} interpret this discrepancy as an effect of subtraction of 
background emission in H$\alpha$ and dust, and claim that 
a super-linear slope results when proper subtraction is performed.
At higher surface densities, a steeper relation is suggested by
observations of Ultra Luminous Infra-Red Galaxies (ULIRGs) and Sub-mm
Galaxies (SMGs) \citep{Bouche07}, but recent observations
\citep{Daddi10b,Genzel10,Boquien11} suggest that at $z\sim2$ ULIRGS and
SMGs, on the one side, and non-IR bright and BzK galaxies, on the
other side, trace two parallel molecular SK relations with slope
$\sim1.4$ and separated by nearly one order of magnitude in
normalization (IR-bright galaxies having higher $\Sigma_{\rm sfr}$).

The interpretation of this phenomenological picture is still under
discussion.  Based on observations, the declining molecular fraction
with gas surface density was proposed by \cite{Blitz04,Blitz06} to be
driven by external pressure, i.e. the midplane pressure of gas in
vertical hydrostatic equilibrium in a thin disk.  However, this
relation has large scatter and is as scattered as other relations
with, e.g., disc mass surface density \citep{Leroy08}.  Alternatively,
many authors \citep{Pelupessy06,Krumholz09a,Gnedin10,Dib11} have proposed
models where the molecular fraction is regulated by the equilibrium
balance between production of H$_2$ and destruction of the same
molecule by UV radiation from young stars (an assumption recently
criticized by \citealt{MacLow10}).  Both creation and destruction
channels are regulated by dust abundance, because dust is both a
catalyst and a shield.  As a consequence, the molecular fraction is
predicted to be driven by gas surface density (or column density) and
modulated by gas metallicity.  Therefore, the scaling of molecular fraction
with gas surface density, or equivalently the saturation value of
$\Sigma_{HI}$, should be a function of metallicity.  \cite{Fumagalli10} have
recently tested the two assumptions (pressure-driven or gas surface
density-driven molecular fraction) against data on nearby spiral and
dwarf galaxies, and report a marginal preference for the second
hypothesis.

The varying slope of the of the molecular SK, steepening toward high
$\Sigma_{\rm cold}$ from $\sim1.0$ to $\sim1.4-1.7$, has been
interpreted by \cite{Krumholz09b} as an effect of the decoupling of
molecular clouds in normal spiral galaxies from the rest of the
Inter-Stellar Medium (ISM).  These authors argue that 
molecular clouds are known to
have a roughly constant surface density and pressure \citep[and then
  dynamical time, see][]{Solomon87}, so they are not in pressure
equilibrium with the rest of the ISM and their consumption time
$\Sigma_{\rm mol}/\Sigma_{\rm sfr}$ results to be $\sim2$ Gyr
\citep{Bigiel08}, irrespective of the molecular gas surface density
computed on $\sim$kpc scale.  This last quantity is indeed to be
considered as a measure of the filling factor of molecular clouds.
The decoupling breaks at higher gas surface densities, where the ISM
is able to pressurize the molecular clouds so that their dynamical
time scales again with the inverse of the square root of the density
at $\sim$kpc scales.  In this regime the molecular SK takes values
more similar to the canonical 1.4 one.

The evidence of a double standard (or molecular) SK at high redshift
has only been proposed very recently. \cite{Teyssier10} have presented
a hydrodynamic simulation, performed with the AMR RAMSES code
\citep{Teyssier02}, of two merging galaxies resembling the Antennae.
They found that, when the force resolution reaches values as small
as 12pc, the predicted standard SK is boosted and reaches the relation
found by \cite{Daddi10b} to hold for ULIRGs and SMGs.  However, the
authors do not show a simulation, run at the same resolution, of the
equivalent of a BzK galaxy that lies on a relation one order of
magnitude lower.

A second way of expressing a ``star formation law'' is by correlating
the surface density of SFR with $\Sigma_{\rm cold}$ divided by the gas
orbital time $t_{\rm orb}$. This dynamical SK was suggested by
\cite{Wyse89}, \cite{Silk97} and \cite{Elmegreen97} to be more
``fundamental'' than the standard SK, on the basis of the influence
that disc rotation and shearing exert on star-forming clouds.  It has
attracted less attention than the standard SK, and in most cases it
has been reported as equally acceptable from the observational point
of view \citep[e.g.][]{Kennicutt98}.  More recently, \cite{Tan10}
tested against observations of local galaxies the hypothesis of a {\em
  linear} standard SK compared to several other proposals of star
formation ``laws''. He found that data do not favour this linear
hipothesis.  At higher redshift, \cite{Daddi10b} noticed that the
dicothomy in the molecular SK of normal and IR-bright galaxies
disappears when the dynamical SK is considered.

The nature of the SK relation and its 
environment-dependent nature is of fundamental importance in the field
of galaxy formation, because SK-like relations are widely used in
models to predict the amount of stars formed in a star-forming galaxy.
In particular, in many N-body hydrodynamic simulations \citep[see,
  e.g.][]{Katz96} the SFR of gas particles or cells is computed as
$\epsilon \times M_{\rm gas}/\tau_{\rm dyn}$, where $\epsilon$ is an
efficiency and the dynamical time is computed on the average density
of the particle.  In many of the most recent simlations of galaxy formation it is
 assumed that the SFR
obeys by construction a standard SK law with a cut at low density
\citep[e.g.][]{Springel03}.  \cite{Schaye08} showed that this is equivalent to
assuming an effective equation of state for star-forming particles of
the kind $P \propto \rho^{\gamma_{\rm eff}}$.  Only higher-resolution 
simulations, resolving scales below 50 pc, are able to follow the
formation of molecules \citep{Pelupessy06, Pelupessy09, Robertson08,
  Gnedin09} and then to predict molecular fractions and kpc-averaged
SK relations, but the cost of these simulations makes it very hard with
presently available facilities to push even one of them to
$z=0$ in a full cosmological context.

One exception in this sense is given by the {\muppi} model
\citep[MUlti-Phase Particle Integrator;][hereafter paper
  I]{Murante10}, based on a sub-resolution multi-phase treatment of
the star-forming gas particles and developed within the {\sc gadget}
TreePM+SPH code \citep{GADGET2}.  In this model, the star formation
rate is {\em not} forced to follow a SK-like relation, so its
adherence to this law is a prediction of the model.  The details of
the model are described below, but two points are worth mentioning.
First, inspired by the observational result of \cite{Blitz06},
  the molecular fraction of the cold component of a gas particle is
  scaled with the hydrodynamic pressure of the SPH particle.
Second, the multi-phase treatment allows thermal feedback
from SNe to efficiently heat the gas.  In paper I, free parameters
were tuned to reproduce the observed SK relation in the case of an
isolated spiral, Milky Way-like galaxy.  We noticed that the SK
relation traced by an isolated low surface brightness dwarf spiral
galaxy differs from the Milky Way one in the same way as spirals and
dwarfs differ in the data of \cite{Bigiel08}.

In this paper we show the SK relations resulting from a set of
{\muppi} simulations of isolated halos, including the ones used in
paper I.  The value of this analysis is not only to present the
results of a specific model but also to understand how the SK
relations depend on galactic environment when the disc is heated by
feedback and the molecular fraction is scaled with pressure and does
not depend on metallicity.  Section 2 describes the {\muppi} model and
the initial conditions used for our simulations.  Section 3 presents
the resulting SK relations. Interpretation of results requires an
assessment of the vertical structure of simulated discs, presented in
Section 3.2. Section 3.3 is devoted to a discussion of the dynamical
SK relation.  Section 4 gives a discussion of present results in
comparison with available literature, while Section 5 presents the
conclusions.

\begin{figure*}
\centering{
\includegraphics[width=0.32\linewidth]{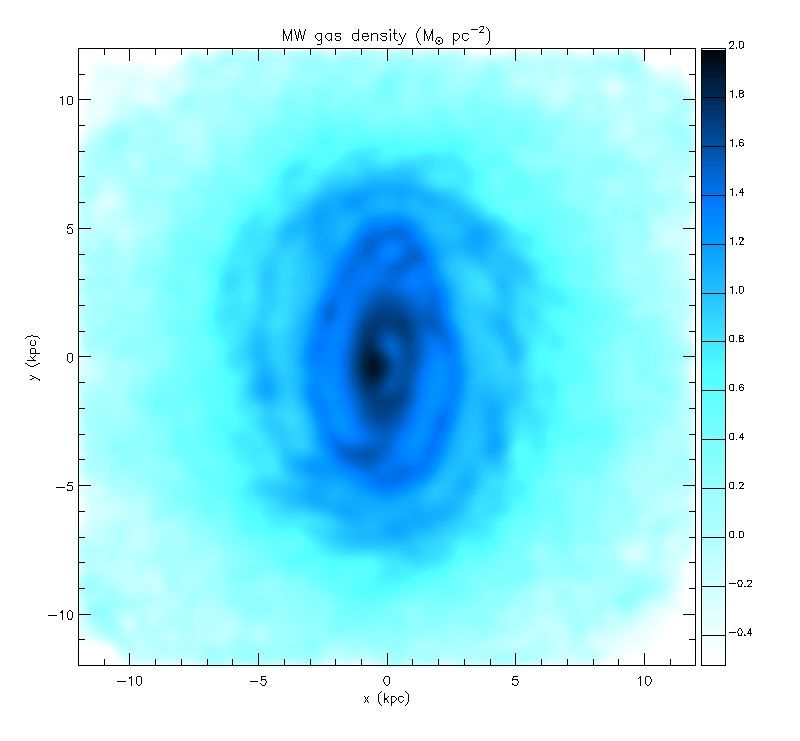}
\includegraphics[width=0.32\linewidth]{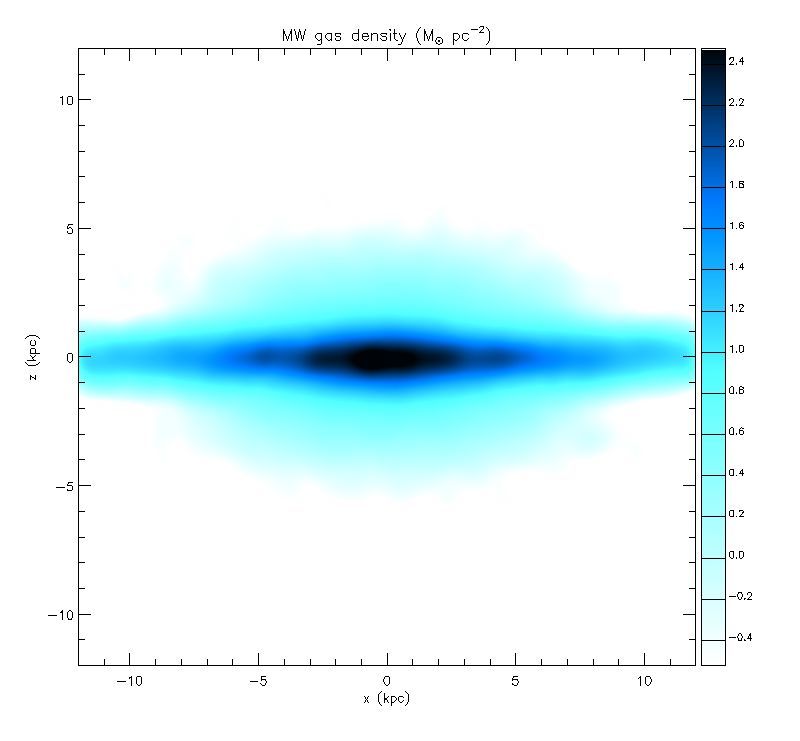}
\includegraphics[width=0.32\linewidth]{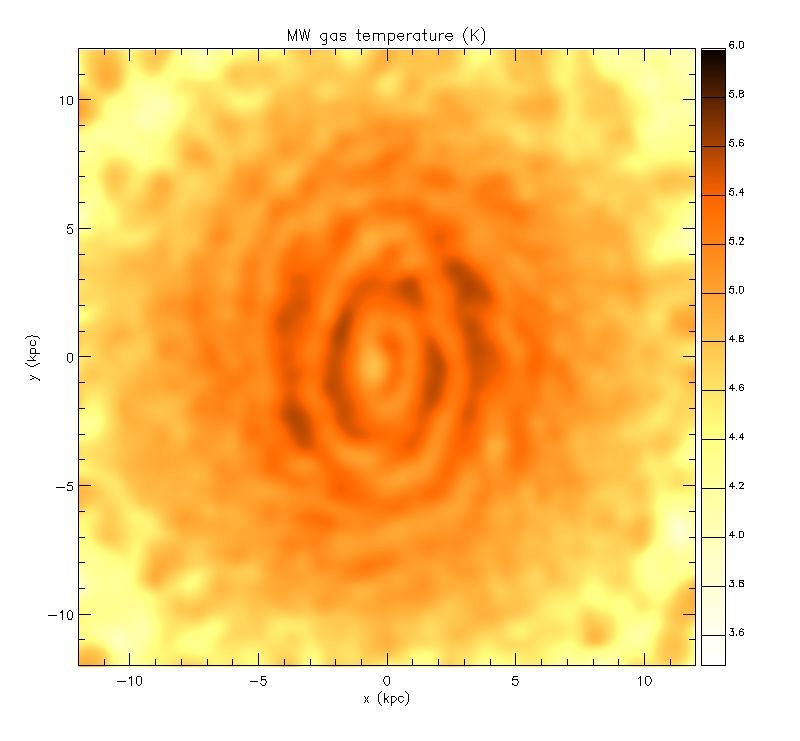}
}
\centering{
\includegraphics[width=0.32\linewidth]{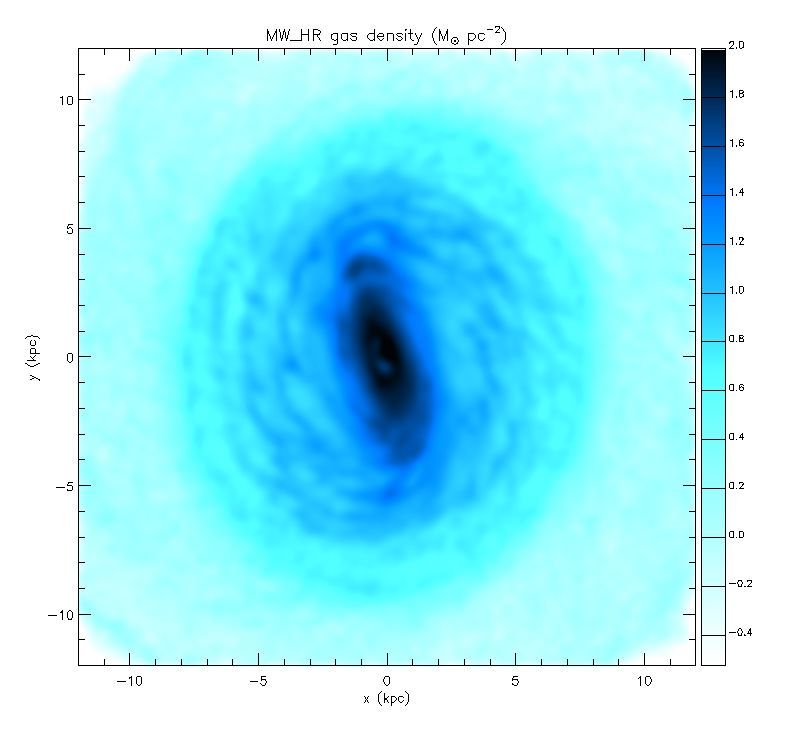}
\includegraphics[width=0.32\linewidth]{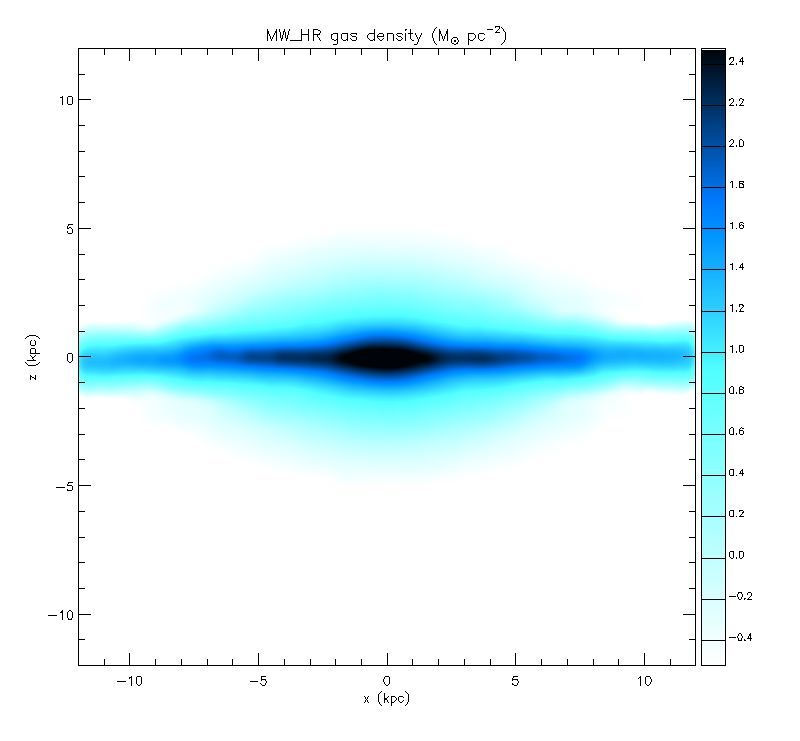}
\includegraphics[width=0.32\linewidth]{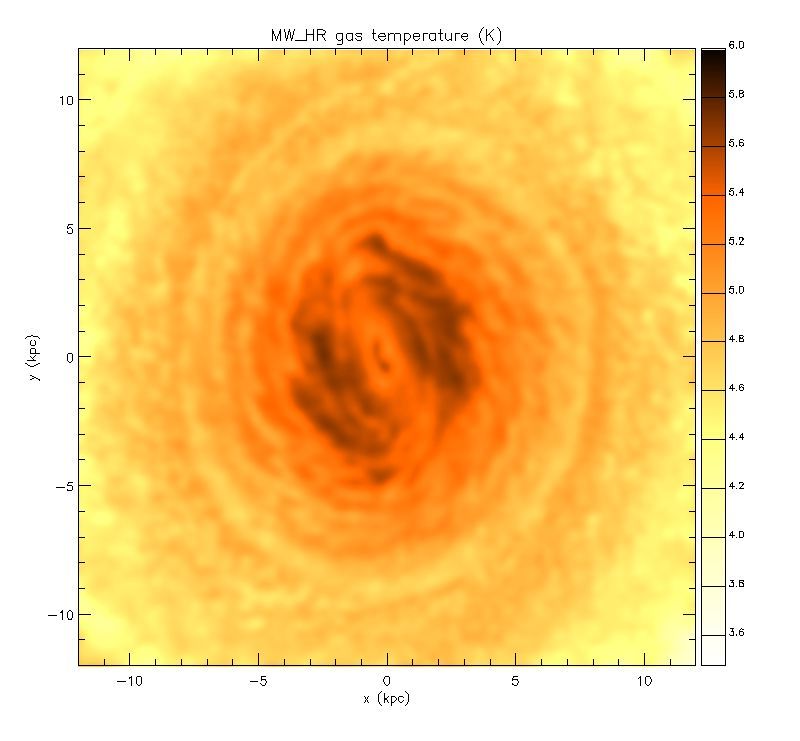}
}
\centering{
\includegraphics[width=0.32\linewidth]{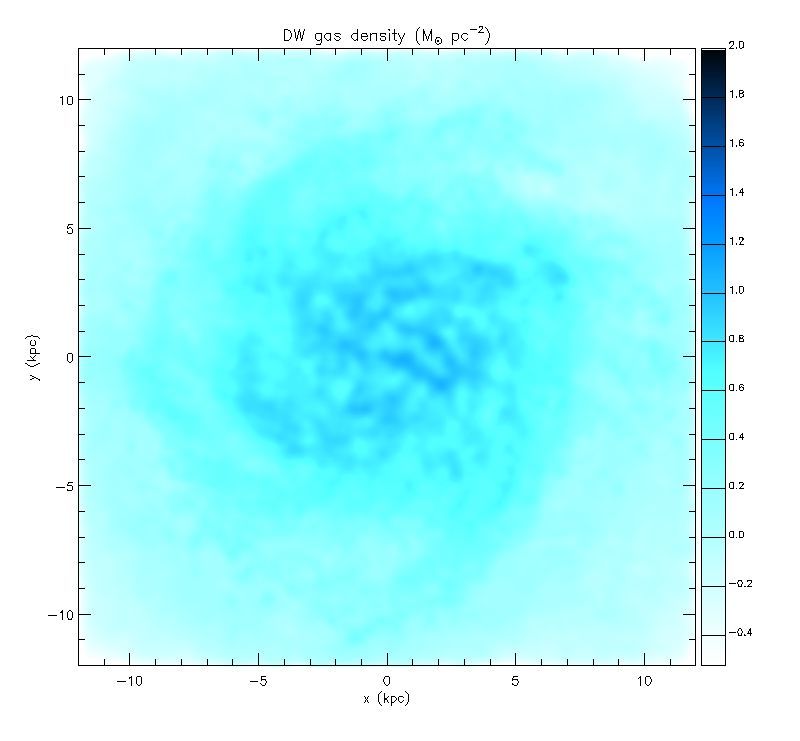}
\includegraphics[width=0.32\linewidth]{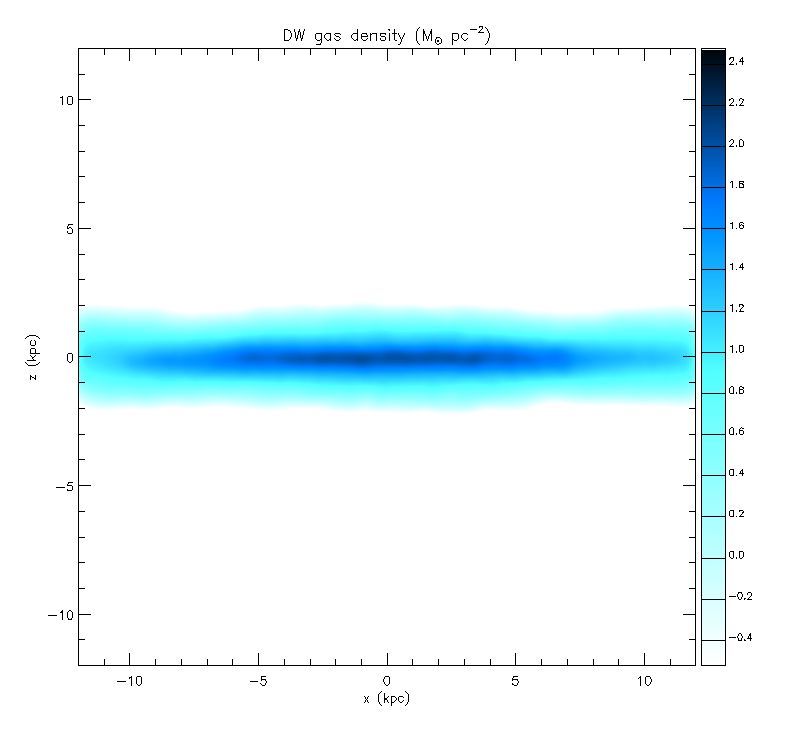}
\includegraphics[width=0.32\linewidth]{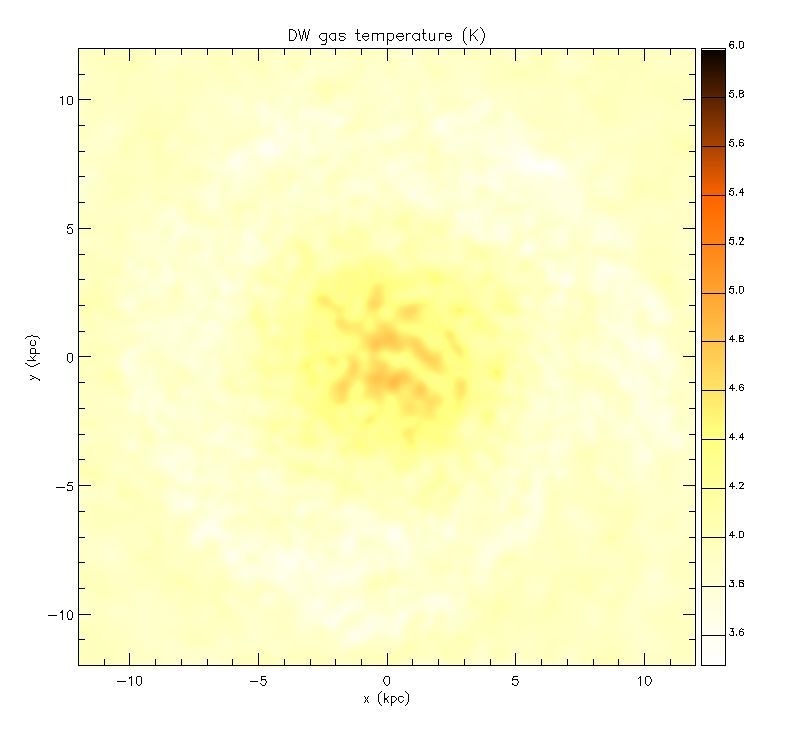}
}
\centering{
\includegraphics[width=0.32\linewidth]{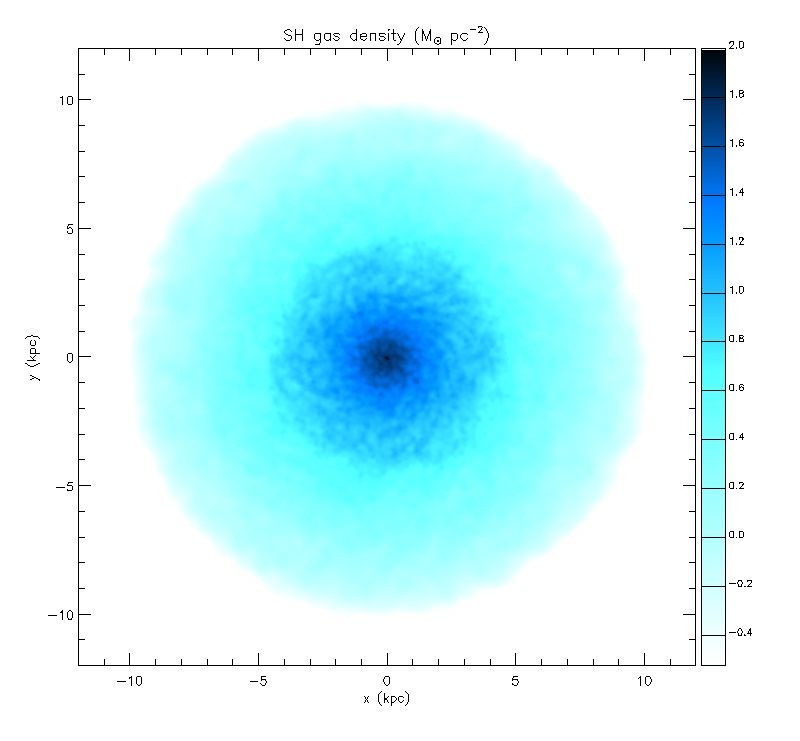}
\includegraphics[width=0.32\linewidth]{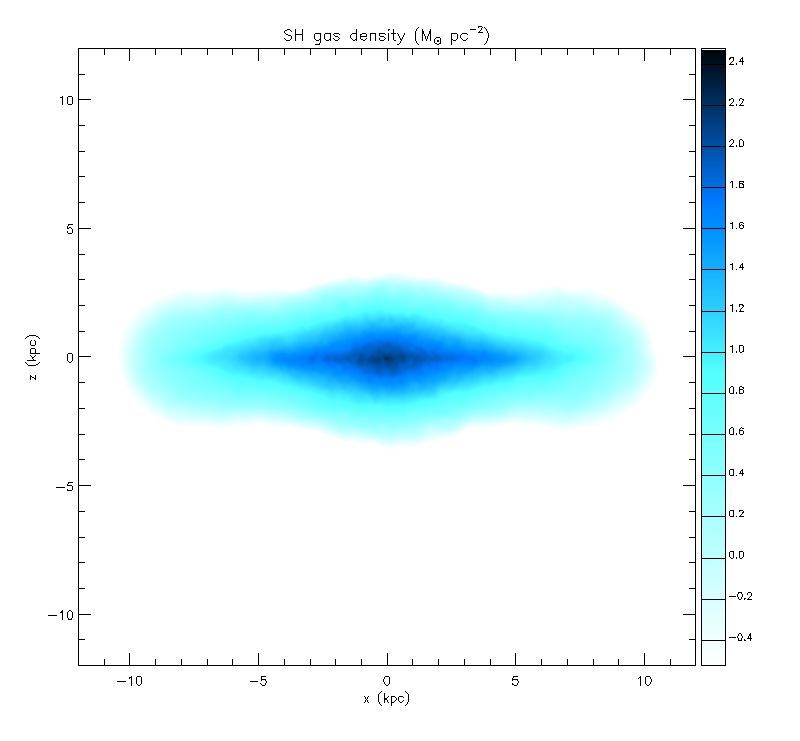}
\includegraphics[width=0.32\linewidth]{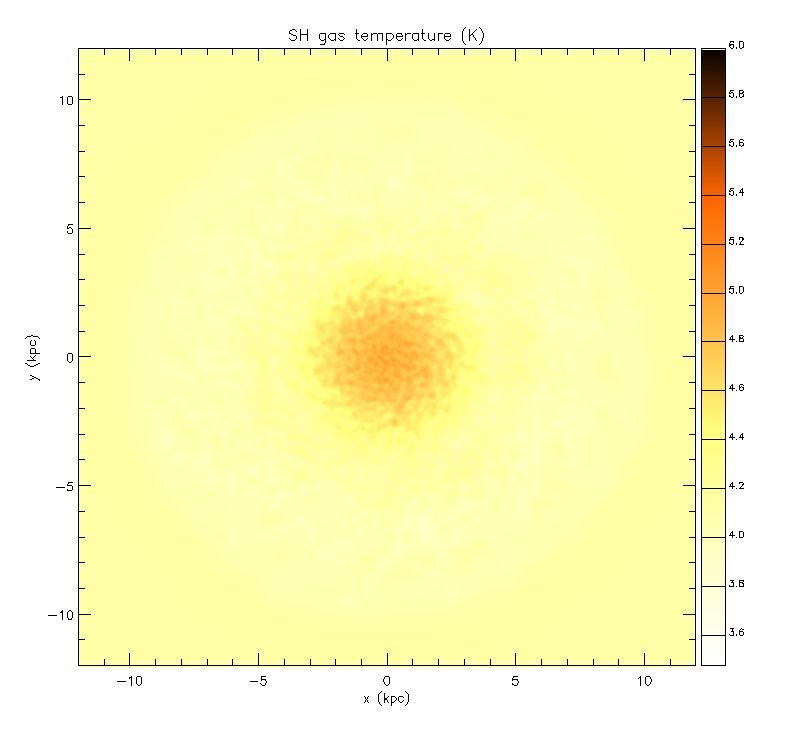}
}
\caption{Gas surface density (left column: face-on; middle column: edge-on)
  and temperature (right column) maps of simulated galaxies. Color
  coding is reported in the color bars, numbers refer to the Log of
  surface density or temperature.  The galaxies shown are, from
  top to bottom, MW, MW\_HR, DW and SH.}
\label{fig:maps}
\end{figure*}

\section{Simulations}
\label{section:simulations}

\begin{table*}
  \caption{
    Basic characteristics of simulated galaxies.
    Column 1: Simulation name.
    Column 2: Gravitational Plummer-equivalent (P-e) softening for gas particles.
    Column 3: Mass of DM halo.
    Column 4: Mass of DM particle.
    Column 5: Stellar mass.
    Column 6: Mass of star particle.
    Column 7: Half-mass radius of stars.
    Column 8: Cold gas mass.
    Column 9: Initial mass of gas particle (before spawning stars).
    Column 10: Half-mass radius of cold gas.
    Column 11: Gas fraction. 
    Notes (1): for MW, MW\_HR and DW stellar masses include both the
    disc and bulge (only MW and MW\_HR) stars present in the initial
    conditions and the newly formed stars, which are a minority. (2):
    in the above cases we report the stellar mass particle of old
    stars, because new stars (whose particle mass is $m_{\rm gas}/4$) 
    give a negligible contribution to the
    disc. (3): The DM halo in the SH simulation is static.
  }
\begin{tabular}{l|c|cc|ccc|ccc|c}
\hline\hline 

    Name & softening & $M_{\rm dm}$ & $m_{\rm dm}$ &$M_\star^{(1)}$&$m_\star^{(2)}$& $R_\star$ & $M_{\rm cold}$ & $m_{\rm gas}$ & $R_{\rm cold}$ & gas   \\
         &   (kpc)   &  (M$_\odot$) & (M$_\odot$)  & (M$_\odot$)   & (M$_\odot$)   &  (kpc)    & (M$_\odot$)    & (M$_\odot$)   &     (kpc)      & fraction \\

\hline

MW       & 0.69 & $9.4\cdot10^{11}$ & $3.5\cdot10^6$ & $4.2\cdot10^{10}$ & $1.3\cdot10^6$ & $4.8$ & $3.3\cdot10^9$ & $7.4\cdot10^4$ & $5.6$ & $7.3$\% \\
MW\_HR   & 0.41 & $9.4\cdot10^{11}$ & $6.9\cdot10^5$ & $4.2\cdot10^{10}$ & $2.6\cdot10^5$ & $4.4$ & $3.2\cdot10^9$ & $1.5\cdot10^4$ & $5.4$ & $7.1$\% \\
DW       & 0.42 & $1.6\cdot10^{11}$ & $8.1\cdot10^5$ & $7.8\cdot10^9$    & $1.6\cdot10^5$ & $8.5$ & $1.9\cdot10^9$ & $3.9\cdot10^4$ & $8.3$ & $20$\% \\
SH       & 0.042& $1.4\cdot10^{10}$ & $-^{(3)}$       & $1.4\cdot10^7$    & $2.2\cdot10^3$ & $0.77$& $1.4\cdot10^9$ & $8.7\cdot10^3$ & $5.2$ & $99$\% \\

\hline
\end{tabular}
\label{table:runs}
\end{table*}

The main properties of the simulated galaxies used in this paper are
reported in Table~1.

The first set of initial conditions is the one used in paper I, and
were generated following the procedure described in \cite{GADGET2}.
They are near-equilibrium distributions of particles consisting of a
rotationally supported disc of gas and stars and a dark matter halo.
For MW and MW\_HR a stellar bulge component is included. Bulge and
halo components are modeled as spheres with \cite{Hernquist90}
profiles, while the gaseous and stellar discs are modeled with
exponential surface density profiles.  To start from a relaxed and
stable configuration, we first evolve the galaxy models for 10
dynamical times with non-radiative hydrodynamics.  We then use the
configurations evolved after 4 dynamical times as initial conditions
for our simulations.  To test the effect of resolution on the SK relations,
the MW galaxy is used also at a higher resolution (MW\_HR).

The second set of initial conditions has been used by
\cite{Springel03} for a resolution test of their star formation code.
Gas particles are embedded in a $1.39\times 10^{10}$ M$_\odot$ static
NFW \citep{Navarro96} halo ($10^{10}$ $h^{-1}$ M$_\odot$ with
$h=0.72$), and rotate at a speed corresponding to a spin parameter of
$\lambda=0.1$, radially distributed following \cite{Bullock01}.  Gas
is initially in virial equilibrium with the halo.  When cooling is
switched on, gas particles slowly coalesce into a rotating disc.  With
this simple setting it is possible to use very high force resolution,
so that the vertical structure of the disc is well resolved.  These
initial conditions are available at four resolutions.  We show results
for only the highest one.  We checked that results are very stable
when the resolution is degraded.

In all cases we forced the SPH smoothing length of gas particles not to drop
below $1/2$ of the Plummer-equivalent softening.

\subsection{The model}
\label{section:model}

{\muppi} has been developed within a non-public version of the {\sc
  gadget}2 Tree-PM+SPH code \citep{GADGET2} that includes an
entropy-conserving formulation of SPH \citep{Springel02}.  It has
already been ported into the more efficient {\gtre} code.  
All details are given in \cite{Murante10}, while we only describe here
some relevant features of the {\muppi} algorithm.

This model is inspired by the multi-phase analytic model of star
formation and feedback by \cite{Monaco04a}.  A particle enters the
multi-phase regime when its temperature is lower than a threshold (set
to $5\times 10^4$ K) and its density, recast in terms of particle
number density (with a molecular weight of $\mu=0.6$) is higher than
0.01 cm$^{-3}$.  This threshold is an order of magnitude lower than
the commonly used value of $\sim0.1$ cm$^{-3}$, which is typically
tuned to obtain a cut in the standard SK relation at a gas density
$\sim10$ {\surf}.

A multi-phase particle is assumed to be made up of three components:
two gas phases and a stellar phase. The two gas phases are assumed to
be in thermal pressure equilibrium.  As in \cite{Springel03}, the cold
gas phase is assumed to have a temperature $T_c=1000$ K,
while the temperature of
the hot gas phase is set by the particle entropy.  Upon entrance into
the multi-phase regime, all the mass is assumed to reside in the hot
phase; cooling does not lead to a lowering of the temperature but to a
deposition of mass in the cold phase.  A fraction of the cold mass is
assumed to be in the molecular form.  \cite{Blitz06} \citep[see
  also][]{Leroy08} found that the ratio between molecular and $HI$ gas
surface densities correlates with the so-called external pressure,
which is the pressure expected at the galaxy midplane in the case of a
thin disc composed by gas and stars in vertical hydrostatic
equilibrium. This was estimated using a simplified version of the 
expression proposed by 
\cite{Elmegreen89}:

\be
P_{\rm ext} \simeq \frac{\pi}{2} G \Sigma_{\rm cold} \left( \Sigma_{\rm cold} 
+ R \Sigma_\star \right)\, .
\label{eq:Pext} \ee

\noindent
Here $R=\sigma_{\rm cold}/\sigma_\star$ is the ration between the
vertical r.m.s. velocity dispersions of cold gas and stars (here
$\sigma$ denotes velocity dispersion while $\Sigma$ denotes surface
density).  The exponent $\alpha$ of the correlation $\Sigma_{\rm
  mol}/\Sigma_{\rm HI}\propto P_{\rm ext}^\alpha$ was found to be
$0.9\pm0.1$. Inspired by this finding, and adopting an exponent
  of 1 for simplicity, we use the following equation to estimate the
molecular fraction:

\be
f_{\rm mol}(P) = \frac{1}{1 + P_0/P}\, .
\label{eq:fmol}
\ee 

\noindent
Here $P$ is the hydrodynamical pressure of the SPH particle (different
from the external pressure used in the observational correlation), and
we adopt $P_0/k=35000$ K cm$^{-3}$ as in \cite{Blitz06}.\footnote{We have
  checked that our simulations produce relations similar to the
  observational one (though with significant scatter) when pressure is
  estimated as in \cite{Blitz06}.}

The three components exchange mass through four mass flows: cooling,
star formation, restoration and evaporation.  Cooling is computed with
a standard cooling function assuming zero metallicity. 
Thermal energy resides in the hot phase, so cooling is computed using
its density, which is much lower than the average one.  Star formation
takes place in the molecular phase, with a consumption time-scale
proportional to the particle dynamical time $t_{\rm dyn}$:

\be
\dot{M}_\star = f_{\rm mol}(P) f_\star M_{\rm cold} / t_{\rm dyn} (n_c)
\label{eq:sfr} \ee

\noindent
Here $f_{\rm mol}$ is the pressure-dependent molecular fraction of
equation~\ref{eq:fmol}, $f_\star=0.02$ is a parameter of star
formation efficiency, determining the fraction of a molecular cloud that
is transformed into stars per dynamical time, and $M_{\rm cold}$ the
mass of the cold phase in the particle.  The dynamical time $t_{\rm
  dyn}$ is computed on the cold phase as soon as this collects 90 per
cent of the total gas mass, and is frozen until the particle exits
from the multi-phase regime (see paper I for a detailed discussion of
this hypothesis).  A thing is worth noticing in equation~\ref{eq:sfr}:
because of the hypothesis of pressure equilibrium and constant
temperature of the cold phase, $n_c$ is proportional to pressure $P$;
at the same time, the fraction of gas mass in the hot phase is always
very low, so $f_{\rm mol}$ is very similar to the fraction of {\em
  total} gas in molecular form.  As a consequence, the particle star
formation rate is primarily regulated by gas pressure (with the
complication that $t_{\rm dyn}$ is computed at the beginning of a star
formation cycle and then kept frozen, while $f_{\rm mol}$ is computed
at each time-step).

The SFR term of equation~\ref{eq:sfr} deposits a fraction $(1-f_{\rm
  re})$ of the transformed mass into the stellar component, while a
fraction $f_{\rm re}$, restored from massive stars in an Instantaneous
Recycling Approximation (IRA), is given back to the hot phase. The
formed stellar component is accumulated within the particle, and
contributes to its inertia but not to its gas mass in all SPH
computations.  The evaporation rate is assumed to be due to the
destruction of molecular clouds and amounts to a fraction $f_{\rm
  ev}=0.1$ of the SFR.

Production of star particles is done according to the stochastic star
formation algorithm of \cite{Springel03} (see paper I for details).
We allow for 4 generations of star particles to be spawned by each
parent gas particle.  Each new star particle
is produced at the expense of the stellar component and, if needed,
of the cold phase.

The three components of a multi-phase particle are of course assumed
to be subject to the same hydrodynamical forces.  To alleviate the
effect of this unphysical assumption, and to mimic the destruction of
a star-forming cloud after a few dynamical times \citep{Monaco04b},
the code forces each particle to leave the multi-phase regime after
two dynamical times $t_{\rm dyn}$, computed as specified above.

One SN is generated each $M_{\star,SN}=120\ {\rm M}_\odot$ of stars
formed, and each SN generates $10^{51}$ erg of energy.  Of the energy
generated in the IRA, a small fraction $f_{\rm fb,i}=0.02$ is given to
the local hot phase to sustain its high temperature, while a fraction
$f_{\rm fb,o}=0.3$ is distributed to the hot phases of neighbour
particles in a 60-degree wide cone anti-aligned with the gas density
gradient.  Energy contributions to particles are weighted by their
distance from the cone axis, to mimic the expansion of SN-driven
blasts along the least resistance path \citep{McKee77}.  The present
version of the code distributes only thermal energy.

The assumption of a molecular fraction regulated by pressure
(equation~\ref{eq:fmol}) is very important because it makes the
evolution of the system intrinsically runaway: star formation
generates SNe, energy feedback from SNe pressurizes the hot phase, the
increase in pressure leads to an increase in molecular fraction and
thus to an increase in SFR.  The runaway halts when the molecular
fraction saturates to unity.  However, the dynamical response of the
pressurized particle is able to limit this runaway through the
expansion work done on neighbours.  This intrinsic runaway behaviour,
together with the long cooling times, are the main reasons for our
efficient thermal feedback.

\begin{figure}
\centering{\includegraphics[width=0.90\linewidth]{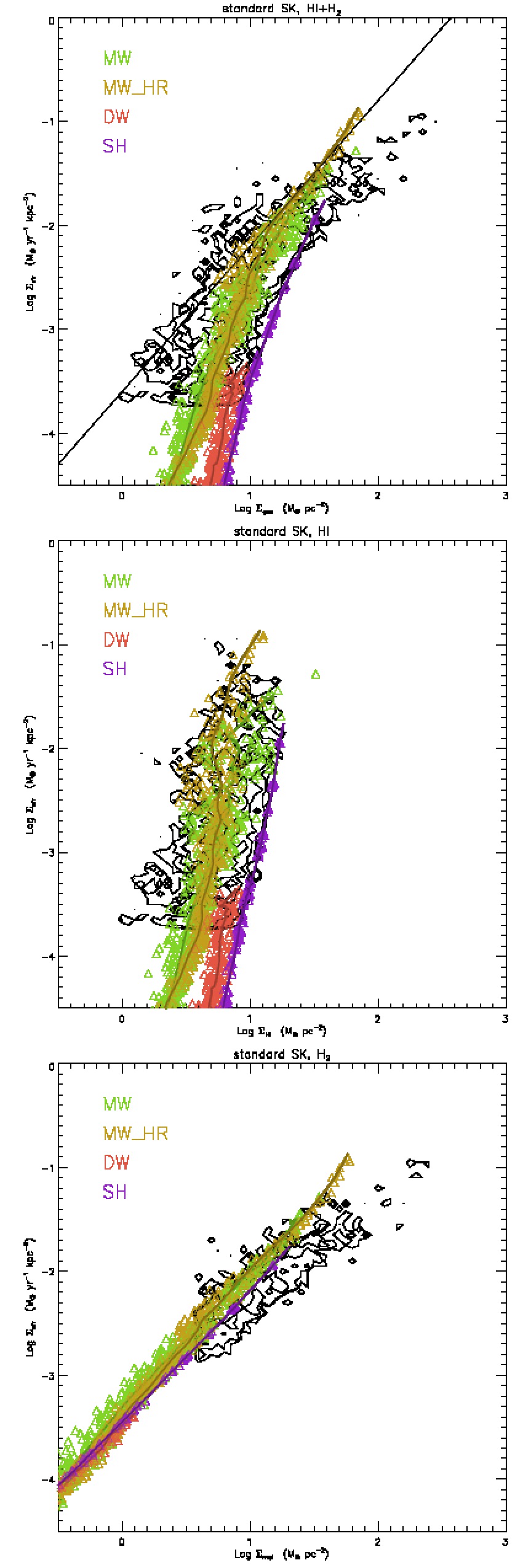}}
\caption{SK relations for simulated galaxies. The black contour levels
  report the data from Bigiel et al. (1998), binned in bins of size
  0.5 dex; levels correspond to 1, 2, 5 and 10 observational points
  per bin, gas surface densities include helium.  Triangles and thick
  lines give results for the four simulations in 750pc bins and in
  radial profiles in cylindrical coordinates; color coding is given in
  each panel.  The three panels give respectively the standard SK
  relation with all gas, HI gas and molecular gas.}
\label{fig:skone}
\end{figure}

\section{The SK relations in simulated discs}

The first three simulations, the MW (at standard and high resolution)
and DW test cases of paper I, start from already formed disc galaxies
with 10 and 20 per cent gas fractions, and are shown at 0.5 Gyr, after
the initial transients due to the switching on of stellar feedback
have (almost) died out and while gas consumption is still negligible.  The
fourth simulation, the SH halo, forms in an isolated, static halo
filled with rotating gas initially in virial equilibrium; we analyze
the simulation at 0.7 Gyr, i.e. at the peak of its SFR, when the disc
is still largely gas-dominated.  In all cases, conclusions are
unchanged when simulations are considered at other times.  We show in
figure~\ref{fig:maps} face-on and edge-on maps of gas surface
densities and temperatures for the four simulations. It is interesting
to notice that in the regions interested by star formation the discs
(most of the MW and MW\_HR discs and the inner few kpc of SH) are relatively
hot and surrounded by thick coronae of gas heated by feedback and
circulating above or below the disc in a galactic fountain. The effect
of star formation on the DW galaxy is much less evident.

\subsection{The standard, HI and molecular SK relations}
\label{section:standard}

Figure~\ref{fig:skone} shows the standard,
HI and molecular
SK relations of the four simulations, compared with the data of
\cite{Bigiel08} for normal spiral galaxies.
Gas surface densities are always meant to include contribution from
helium.  In the upper panel the thin line represents the fit proposed by
\cite{Kennicutt98}.  Simulations have been processed as follows.  Our
analyses are restricted to cold gas; in this paper by cold gas we mean
the cold phase of multi-phase particles plus all single-phase
particles colder than $10^5$ K.  The molecular gas surface density is
computed using the molecular fraction of equation~\ref{eq:fmol},
applied only to multi-phase particles; HI gas is just cold minus
molecular gas.  A galaxy frame is defined by the inertia tensor of
stars and cold gas, the z-axis corresponding to the largest
eigenvalue.  The angular momentum of the same particles is always
found to be at most a few degrees off the z-axis.  Then, radial
surface densities (in cylindrical coordinates) of (cold, HI,
molecular) gas and SFR are computed; these are reported in the figure
as colored thick lines.  The same quantities are computed on a square
grid in the x-y plane, with bin size of 750 pc, as in the
\cite{Bigiel08} paper; these are reported as triangles, with the same
color as the corresponding thick lines.

The following points are apparent from the figure: (i) the simulations
trace different SK relations when the total or HI gas are used.  (ii)
These different relations correspond to different transitions from HI-
to molecular-dominated gas, or equivalently to different saturation
values of $\Sigma_{HI}$.  (iii) The differences go in the direction
highlighted by \cite{Bigiel08}, \cite{Bolatto09,Bolatto11} and
\cite{Bigiel10} of a higher saturation $\Sigma_{HI}$ in dwarf
galaxies.  (iv) The molecular SK relation is much tighter; also, it is
steeper than the \cite{Bigiel08} one and consistent with \cite{Liu11}.
(v) The scatter around the SK relation is generally smaller than the
data, especially for the SH simulation.  (vi) As for the MW and MW\_HR
simulations, the SK relation is rather stable with resolution; this is
confirmed by applying the same analysis to the SH halo at lower
resolution.

\begin{figure}
\centering{\includegraphics[width=0.95\linewidth]{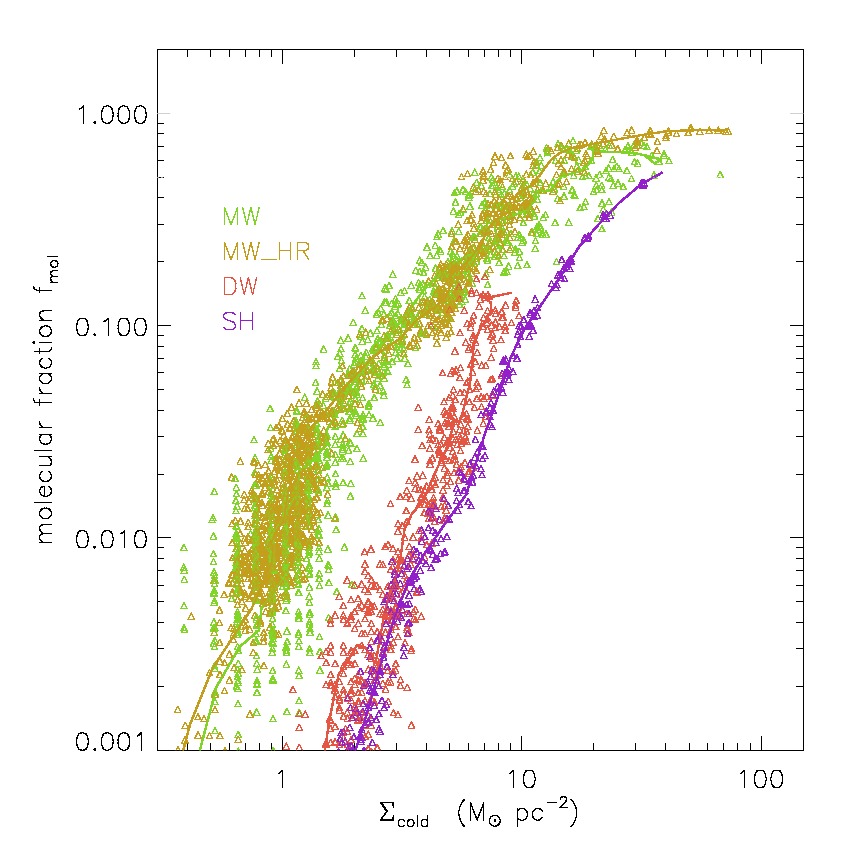}}
\caption{Molecular fraction as a function of cold gas surface density
  for the four simulations.  The dotted line marks the
  $1/2$ value, where HI and molecular gas surface densities are
  equal.}
\label{fig:R}
\end{figure}

To better quantify the difference between the simulations, in
figure~\ref{fig:R} we show the relation between surface density
and molecular fraction $f_{\rm mol}$, a quantity that is directly
comparable with data. The $\Sigma_{\rm cold}$ value at which the
molecular fraction is $1/2$ ranges from 10 {\surf} of MW to 35 {\surf}
of SH.  This is in agreement with the measures of the Large Magellanic
Cloud, though more extreme cases like the Small
Magellanic Cloud are not recovered (\citealt{Bolatto09,Bolatto11}; \citealt[see also references
  in][]{Fumagalli10}).

\begin{figure*}
\centering{\includegraphics[width=0.95\linewidth]{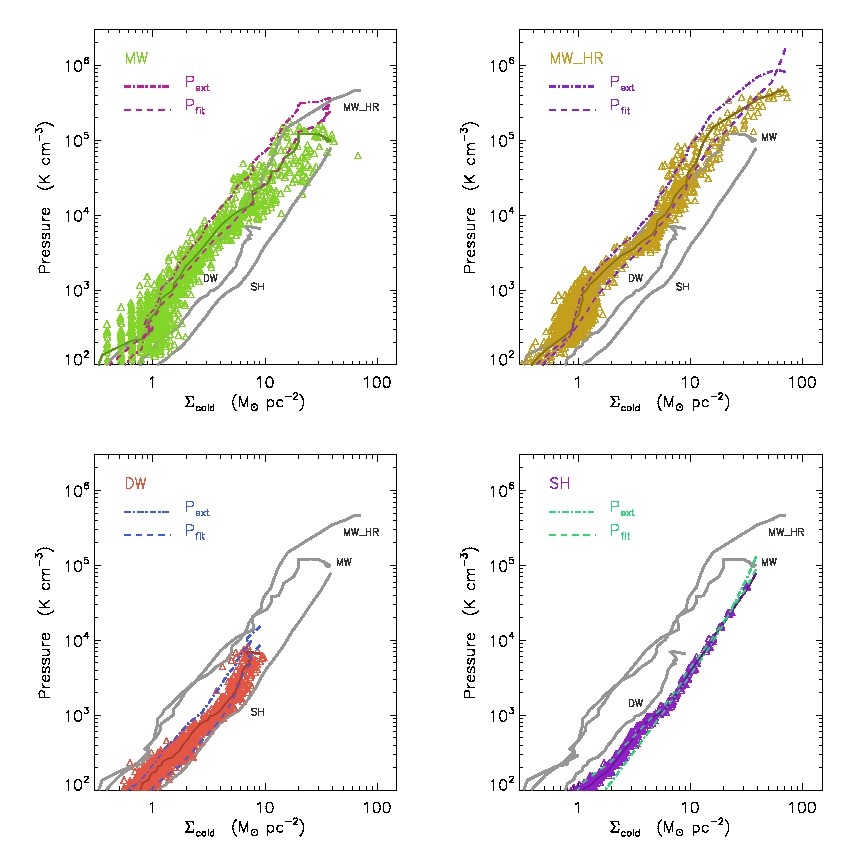}}
\caption{
Relation between total cold gas surface density and hydrodynamical pressure
  in simulations.
  Triangles denote the relation in 750pc bins, while
  darker continuous lines of the same color give the radial averages.  
  The bright
  dot-dashed and dashed lines give radial estimates based respectively
  on hydrostatic equilibrium (equation~\ref{eq:Pext}) and on
  equation~\ref{eq:Pfit}.  Each panel contains the results from one
  simulation, while only the radial averages of hydrodynamical pressure from
  the other simulations are reported in each panel as grey lines.}
\label{fig:P}
\end{figure*}

As pointed out by \cite{Fumagalli10}, if the molecular SK is the
``fundamental'' relation then a pressure law for the molecular
fraction results in an environment-dependent standard SK.  This is
true because, as evident in equation~\ref{eq:Pext} for pressure in the
case of vertical hydrostatic equilibrium, this quantity depends not on
gas surface density alone (the quantity in the x-axis of the standard
SK) but on gas {\em and} stellar surface densities, the latter being
multiplied by the ratio $R$ of gas and star vertical velocity
dispersions \citep[see also][]{Shi11}.  In other words, the cut in the standard SK is determined
by the (velocity-weighted) gas fraction.  So our result is easy to
interpret as long as hydrodynamic pressure of our discs scales with
gas fraction in a similar way as the external one.  It is worth
noticing at this point that our simulated galaxies have gas fractions
ranging from low to very high values, so they span most of the
interesting range for this quantity.

However, the validity of equation~\ref{eq:Pext}, obtained in the case
of gas in vertical hydrostatical equilibrium, must be checked in our
simulations where energy is continually injected in discs.  This is
done in next sub-section.

\subsection{Vertical structure of simulated discs}
\label{section:vertical}

\begin{figure}
\centering{\includegraphics[width=0.95\linewidth]{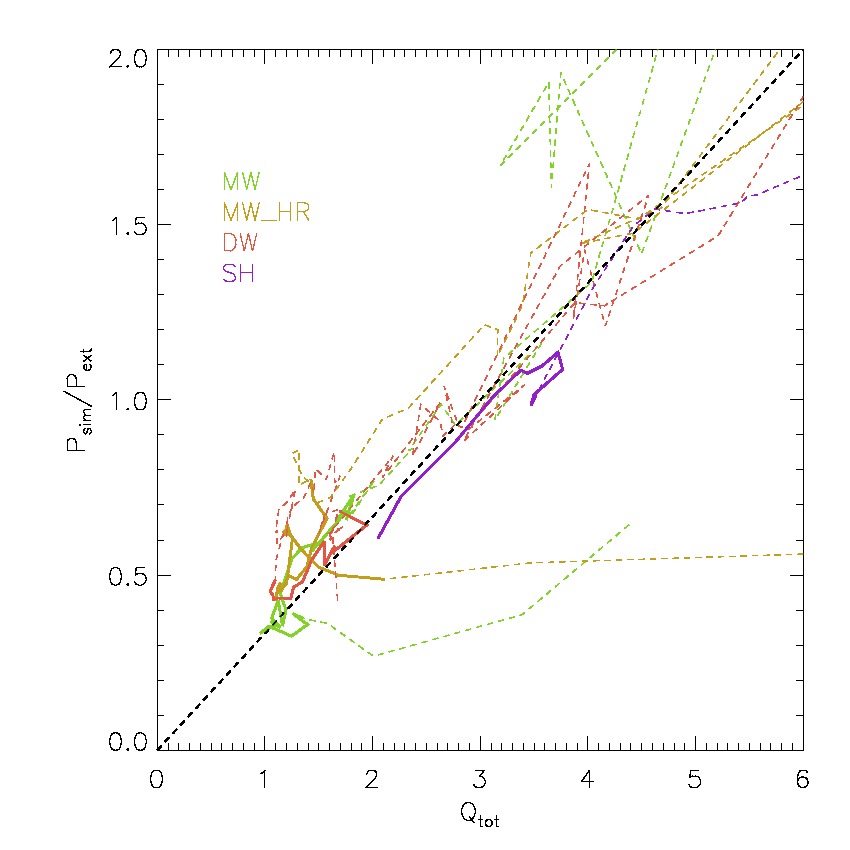}}
\caption{Correlation of the ratio of the radial average of $P_{\rm
    sim}/P_{\rm ext}$ with Toomre parameter $Q_{\rm tot}$ for the four
  simulations.  Color coding is reported in the panel; thick
  continuous lines correspond to that part of the galaxy beyond two
  softening lengths from the center and with significant SFR, thin
  dashed lines refer to the rest of the galaxy.  The dashed thick line
  has slope $1/3$.}
\label{fig:corr_Q}
\end{figure}

In figure~\ref{fig:P} we report, for the four simulated galaxies, the
relation between pressure and $\Sigma_{\rm cold}$ as found in the
simulations.  We show in each panel the hydrodynamical pressure found
in simulations, $P_{\rm sim}$, averaged in bins of the x-y grid as
colored triangles, and the radial profile of the average of the same
quantity as a line with a darker color.  The external pressure $P_{\rm
  ext}$ is computed using equation~\ref{eq:Pext} from the radial
profiles (colored dot-dashed lines).  Moreover, to ease the comparison,
each panel reports, as grey thick lines, the radial $\Sigma_{\rm
  cold}-P_{\rm sim}$ relations from the other simulations.  To obtain
$P_{\rm ext}$, vertical velocity dispersions $\langle v_z^2 \rangle$
of cold gas and star particles have been computed in the galaxy frame.
For the stars, this quantity is equated to $\sigma_\star^2$, while for
the gas we use:

\be
\sigma_{\rm cold}^2 = \langle v_z^2 \rangle + c_s^2 \, ,
\label{eq:scold}
\ee

\noindent
where $c_s$ is the gas sound speed, computed on the average particle
temperature and density.  Figure~\ref{fig:P} shows that the relation
between $P_{\rm sim}$ and $\Sigma_{\rm cold}$ varies in a very similar
way as that between $P_{\rm ext}$ and $\Sigma_{\rm cold}$.  This
confirms the validity of our interpretation for the environmental
dependence of the standard SK.  However, external and simulated
pressures show significant dicrepancies that are worth addressing.

\begin{figure}
\centering{\includegraphics[width=0.95\linewidth]{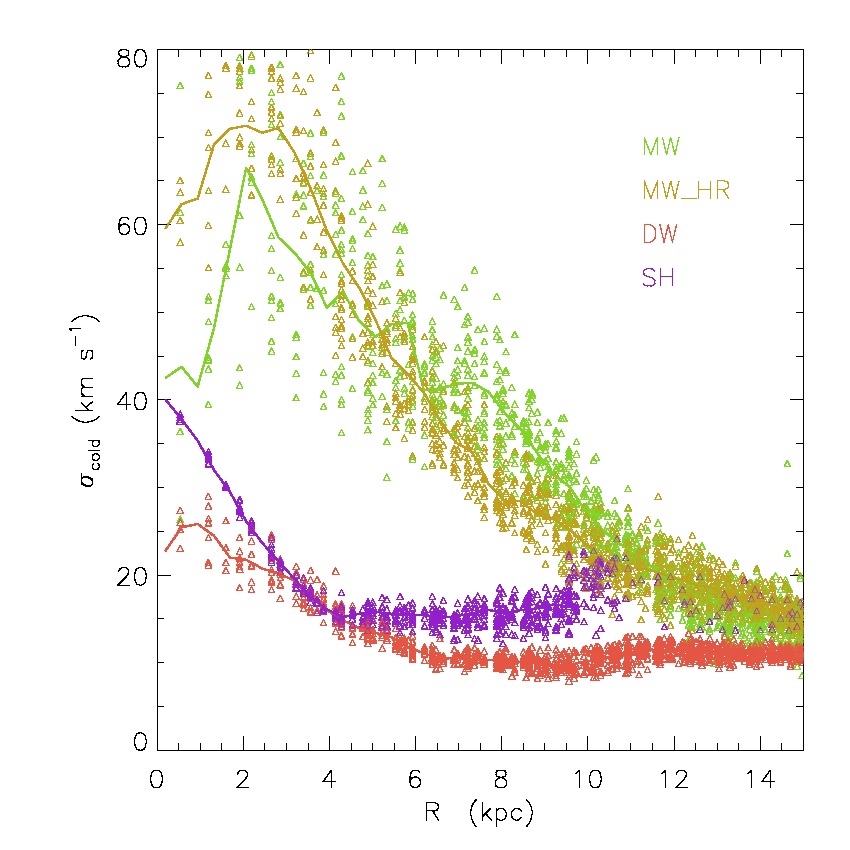}}
\caption{Gas vertical velocity dispersions, including thermal and
  kinetic energy (\ref{eq:scold}), for the four simulations.  Color
  coding is reported in the panel.}
\label{fig:sigma}
\end{figure}

These discrepancies may be related to the fact that we are comparing
the expected midplane pressure with the average one over the whole
disc height.  While for barely resolved discs the two quantities cannot
differ much, the SH simulation allows us to test to what extent
midplane and average pressures are comparable. As a matter of fact,
particles in the midplane show a broad range of pressures, because the
accumulation of cold mass at the beginning of a star formation cycle
causes a depressurization by one order of magnitude, while successive
star formation re-pressurizes the star-forming particles.  As a
result, pressure does not show a smooth decrease with height on the
disc, and the (mass-weighted) average of pressure over the disc height
is always very similar to the midplane one. A similar trend was found
by \cite{Tasker08}, using a much better resolution, when energy
feedback from SNe is considered.

We noticed that in our simulations the ratio of true pressure $P_{\rm
  sim}$ and $P_{\rm ext}$, computed the first in radial bins and the second from radial
profiles, correlates well with the Toomre parameter $Q$ of the disc,
computed at the same radius.  For a disc made of a single component
with surface density $\Sigma$ and velocity dispersion $\sigma$, we have $Q(r)
= \sigma \kappa / \pi G \Sigma$, where $\kappa=V(r)\sqrt{2 + 2
  d\ln V/d\ln r}/r )$ is the epicyclic
frequency and $V(r)$ the rotation curve.  Being our discs composed by stars and gas, we compute the
disc total Toomre parameter $Q_{\rm tot}$ by using the simple
approximation of \cite{Wang94}, that is correct for our discs
characterized by relatively high velocity dispersions \citep{Romeo11}:

\be Q_{\rm tot} \simeq \left( \frac{1}{Q_{\rm cold}} +
\frac{1}{Q_\star} 
\right)^{-1} 
= \frac{\sigma_{\rm cold} \kappa}{\pi G (\Sigma_{\rm cold} + R \Sigma_\star)}\, .
\label{eq:Q} \ee

\noindent
It is easy to show that $Q_{\rm tot}=\sigma_{\rm cold} \kappa
\Sigma_{\rm cold} / 2 P_{\rm ext}$.

\begin{figure}
\centering{\includegraphics[width=0.95\linewidth]{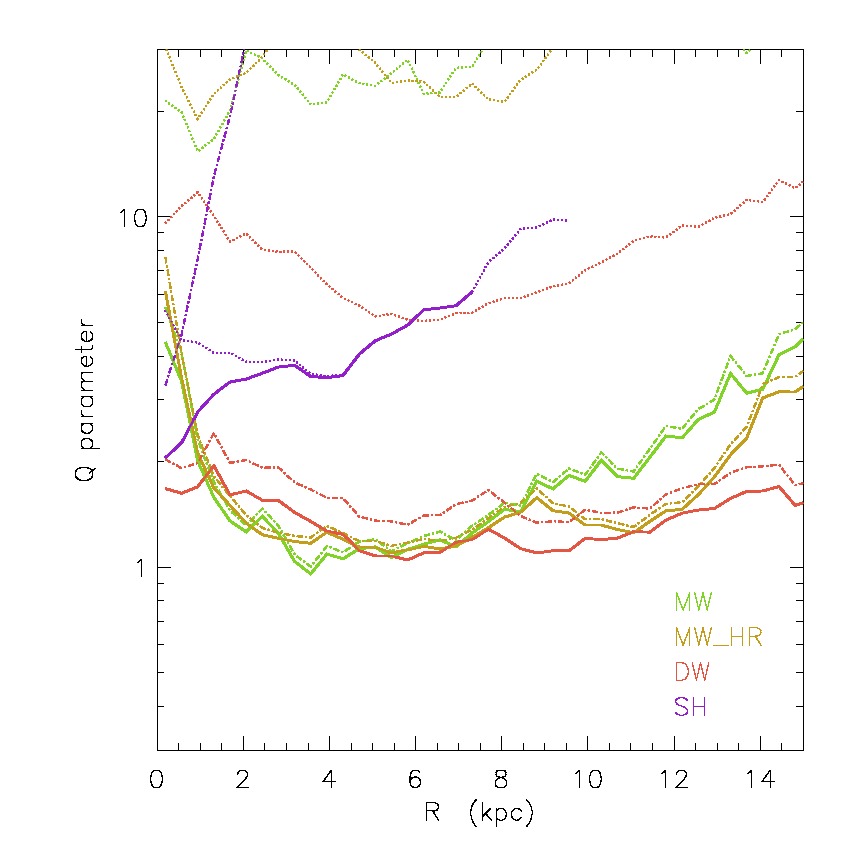}}
\caption{Toomre Q parameters for cold gas (dotted lines), stars (dashed
  lines) and total (continuous lines).  Color coding is reported in
  the panel.}
\label{fig:Q}
\end{figure}

Figure~\ref{fig:corr_Q} shows for our four simulations and at all
radii the relation between $Q_{\rm tot}$ and the ratio $P_{\rm
  sim}/P_{\rm ext}$.  The most interesting regions with $\Sigma_{\rm
  sfr}>10^{-5}$ {\ssfr} and radius larger than two softening lengths
have been highlighted as thick continuous lines.  The two quantities
show a linear  correlation that is
well fit by $P_{\rm sim} / P_{\rm ext} = Q_{\rm tot} / 3$; the round
number 3 adapts well to the SH simulation where the vertical structure
of the disc is fully resolved.  Then, a much better fit to the $P_{\rm
  sim} -\Sigma_{\rm cold}$ relation is given by:

\be
P_{\rm fit} = P_{\rm ext} \times \frac{Q_{\rm tot}}{3} 
 = \frac{1}{6} \Sigma_{\rm cold} \sigma_{\rm cold} \kappa\, .
\label{eq:Pfit} \ee

\noindent
The second equality is obtained using equations~\ref{eq:Pext} and
\ref{eq:Q}.  $P_{\rm fit}$ is shown in figure~\ref{fig:P} as dashed
lines; it gives a much better approximation to the $P_{\rm sim} -
\Sigma_{\rm cold}$ relation, with some discrepancy for the DW galaxy
at $\Sigma_{\rm cold}<5$ {\surf}, where $\Sigma_{\rm sfr}$ is anyway
very low.  We checked the validity of $P_{\rm fit}$ by comparing it to
the pressure profile of many versions of our galaxies, obtained in our
tests by varying model parameters and assumptions.  In particular, we
tested changes in the computation of dynamical time used in
equation~\ref{eq:sfr} in several ways, e.g. by equating it to the
average one at entrance into multi-phase or by recomputing it at each
time-step.  We found in all cases that equation~\ref{eq:Pfit} always
gives a good fit to pressure, with significant discrepancies found
only at very low pressure.  Therefore we consider this result to be
independent of details of our sub-resolution model.   Presently,
  we have no analytical explanation of why multiplying the external
  pressure by $Q_{\rm tot}/3$ gives a better fit to the average SPH
  pressure in our simulations.  We interpret equation \ref{eq:Pfit}
as the quasi-equilibrium pressure of a disc with continuous injection
of energy from SNe: hotter discs, with high $Q_{\rm tot}$, are
characterized by higher pressure than $P_{\rm ext}$, while marginally
stable discs have a lower pressure by a factor up to $\sim3$ (a
similar result was found by \citealt{Koyama09}, see section 4 for more
details).

More insight on the vertical structure of the disc can be obtained by
analysing the quantities $\sigma_{\rm cold}$ and $Q_{\rm tot}$, that
enter in the computation of $P_{\rm fit}$.

Figure~\ref{fig:sigma} shows the values of gas vertical velocity
dispersions $\sigma_{\rm cold}$ for the four galaxies, computed both
on the 750 pc grid and in radial profiles.  Clearly the assumption of
a constant velocity dispersion, often made in the literature to
represent this quantity, is a poor approximation of our results.
Moreover, these velocities are much higher than the $\sim5-10$ {\kms}
usually assumed for galaxy discs.\footnote{This quantity includes
  contribution from the thermal energy of the hot phase, so it is not
  directly comparable to observations. We will return on this in section 4.}

The values of the Q parameters for
the four simulations are shown in figure~\ref{fig:Q}.  $Q_{\rm tot}$
was computed as in equation~\ref{eq:Q}, the same parameter was
computed only for gas and stars as $Q_i = \sigma_i \kappa / \pi G
\Sigma_i$ (where $i$ is either cold or $\star$), and its values are
reported in the figure with dotted and dashed lines, respectively.  
The MW and DW discs have $Q_{\rm tot}\sim 1$ in the central,
star-forming regions, while $Q_{\star}$ assumes higher values.  $Q_{\rm cold}$
is very high, $\sim20$; this quantity cannot be directly compared with
observations of cold gas, because the main contribution to
$\sigma_{\rm cold}$ comes from the sound speed, 
that is determined by the thermal
energy of the hot phase of multi-phase particles.  If the sound speed is
neglected, we obtain $Q_{\rm cold}\sim5-10$.  With this caveat,
these values of the
Toomre parameters are in very good agreement with what found by
\cite{Leroy08} in nearby galaxies, where $Q_{\rm tot}$ is just above
unity while $Q_{\star}$ and especially $Q_{\rm cold}$ take up higher
values.  At the same time, the gas-dominated SH disc is much more
stable and kinematically hot.  As a conclusion, while
our stellar-dominated discs tend to regulate to keep 
$Q_{\rm tot}\sim 1$, this is not a general rule.

As a further test, we run our simulations, starting from the same
four sets of initial conditions but using the effective model of
\cite{Springel03}, that is known not to provide efficient thermal
feedback.  We obtained colder discs, with higher $Q_{\rm tot}$
parameters and hydrodynamical pressure not well fit either by $P_{\rm
  ext}$ or by our equation~\ref{eq:Pfit}.  The resulting standard SK
relation is in line with that originally found by \cite{Springel03}: it lies on the
\cite{Kennicutt98} relation above $\sim$5 {\surf}, and cuts below that
threshold, not following the mild steepening found in data.  This
surface density threshold somehow depends on the galaxy: we find it at
3, 4.5 and 6 {\surf} for the MW, DW and SH simulations.  This
dependence is easy to understand: the cut is determined by the imposed
volume density threshold for star formation. Since disc
temperatures do not vary much (especially in these colder discs) the
relation between volume and surface density of gas scales similarly to
that between pressure and surface density.  Of course no prediction
can be made in this case for the HI and molecular SK.  We conclude that
a simple volume density threshold for star formation, used in
conjunction with an ineffective feedback scheme, gives a standard SK
with environmental dependence that goes in the same direction as
the one presented in this paper, but cannot reproduce the data at the
same level of detail.

\subsection{The dynamical SK}
\label{section:origin}

\begin{figure}
\centering{\includegraphics[width=0.95\linewidth]{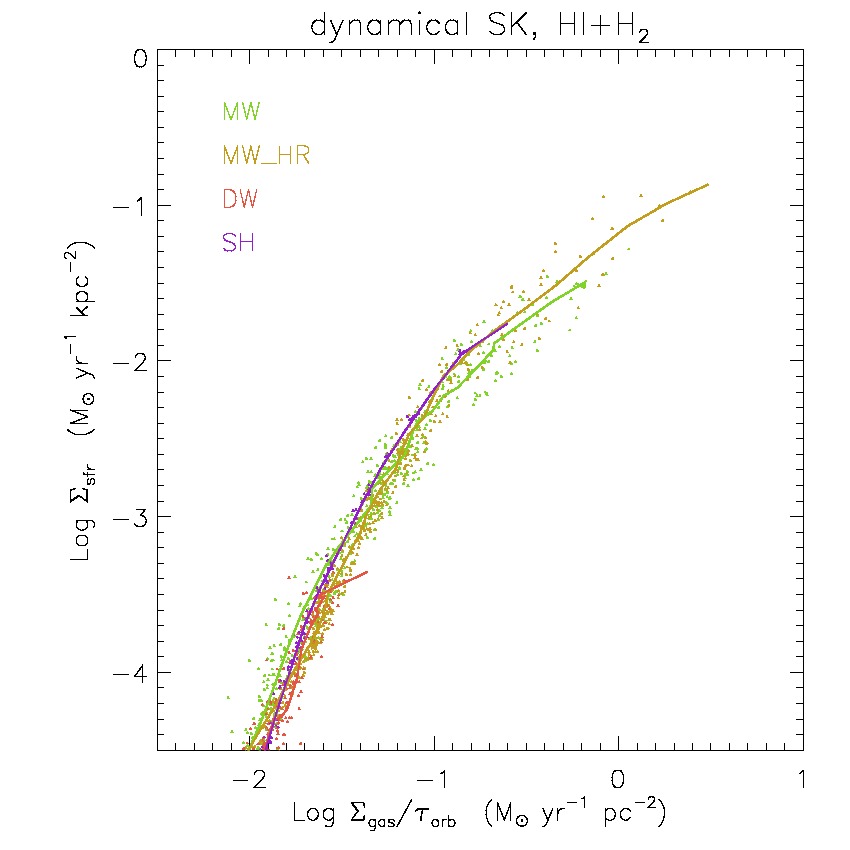}}
\caption{Dynamical SK relations for simulated galaxies. Points and
  thick lines give results for the four simulations in 750pc bins and
  in radial profiles in cylindrical coordinates; color coding is given
  in each panel.  Point size has been decreased to make the agreement
  of the four relations more evident.}
\label{fig:sktwo}
\end{figure}

Figure~\ref{fig:sktwo} shows the dynamical SK of our four simulations.
Unfortunately, a dataset like that of \cite{Bigiel08} is unavailable
for this relation; most observations give global estimates of
galaxies. As a consequence, we do not compare this relation with data
in this paper.

Remarkably, the four simulations, that produced different standard
SK's, now trace a unique, non-linear relation, with a scatter that is even lower
than that in the molecular SK.  
To make this equality more
evident, we decrease the point size of the grid-based relation to let
radial profiles be more visible.

It is easy to show, using equations~\ref{eq:Pext} and \ref{eq:Q}, that a ``universal'' dynamical SK relation, valid
for all galaxies, can be obtained under the following simple
assumptions: (i) the disc is in vertical hydrostatic equilibrium, so
that pressure is well represented by $P_{\rm ext}$; (ii) the disc is
marginally stable, with a $Q_{\rm tot}$ Toomre parameter equal to 1;
(iii) the velocity dispersion of gas $\sigma_{\rm cold}$ is constant
for all galaxies; (vi) the rotation curve is flat so that $\kappa
\simeq \sqrt{2}/\tau_{\rm orb}$; (v) SFR is a function of pressure.
But we have shown above that the first three conditions don't hold for
our discs, so this simple interpretation cannot be valid here.  On the
other hand, we noticed that our simulated discs are rather hot when we
take into account the effective temperature of gas particles, which is
determined by the thermal energy of the hot phase.  In the following
we demonstrate that the dynamical SK can be interpreted as the result
of the balance between energy injection and dissipation, without
assuming that SFR is directly influenced by disc rotation.

Energy is continually injected by SN feedback and lost by radiative
cooling and viscous dissipation.  The equilibrium among these
processes can be illustrated with the help of a simplified approach that
quantifies the energy made available to gas particles, in
both kinetic and thermal form, after
cooling has radiated away a part of it.  
Injection of kinetic and
thermal energy can be expressed as (see also equation~14 of paper
I):

\be
\dot{\Sigma}_{\rm inj} = \epsilon v^2_{\rm sn} \Sigma_{\rm sfr}\, ,
\label{eq:heating}
\ee

\noindent
where the constant $v^2_{\rm sn}=(f_{\rm fb,i}+f_{\rm fb,o})
10^{51}\ {\rm erg}/M_{\star,sn}$ takes account of the energy made
available for feedback by {\muppi} and $\epsilon$ is an efficiency
parameter that quantifies the fraction of injected energy that is {\em
  not} radiated away by cooling.  This injected energy is then
transformed by expansion into kinetic energy and is later dissipated
by viscosity\footnote{
The fact that in our simulations viscosity is the numerical one
included in our SPH code is at this stage immaterial, as long as this
behaves similarly to dissipation of turbulence that would take place
if the resolution were adequate to describe it.
}; this keeps the disc in a quasi-stationary state.  For a disc of
height $H_{\rm eff}$, disc energy will likely be dissipated on one
disc height crossing time, $t_{\rm cross} = H_{\rm eff}/\sigma_{\rm
  cold}$.  Notably, this is the same rate at which turbulence is
dissipated, as long as the driving length of turbulence is the size of
the typical SN-driven bubble \citep[e.g.][]{MacLow03}, which must be
$\sim H_{\rm eff}$ if bubbles die by blowing out of the disc
\citep[see also][]{Monaco04a}.

Let's define the surface density of energy in the disc as:

\be
\Sigma_{\rm E} = \frac{3}{2}\Sigma_{\rm cold}\sigma_{\rm cold}^2\, ,
\label{eq:Se}\ee

\noindent
where we assume equipartition between the three translational degrees
of freedom.  Then the energy dissipation rate is:

\be \dot{\Sigma}_{\rm disp} = \frac{\Sigma_{\rm E}}{t_{\rm cross}}
= \frac{3\Sigma_{\rm cold} \sigma_{\rm cold}^2 }{2t_{\rm
    cross}}\, ,
\label{eq:diss}
\ee

In the definition of effective disc height $H_{\rm eff} = \Sigma_{\rm
  cold}/2\rho_{\rm cold}$ the midplane density can be substituted with
pressure using the equation of state $P=\rho_{\rm cold}\sigma_{\rm
  cold}^2$.  Using equation~\ref{eq:Pfit} for the pressure, it is easy
to show that:

\be
H_{\rm eff} = 3 \frac{\sigma_{\rm cold}}{\kappa}\, .
\label{eq:Heff} \ee

\noindent
It then follows that:

\be
t_{\rm cross} = \frac{3}{\kappa} \simeq \frac{3}{\sqrt{2}} \tau_{\rm orb}\, .
\label{eq:tcross}
\ee

In a stationary system, the effective energy injection will equate
dissipation, so that $\dot{\Sigma}_{\rm inj} = \dot{\Sigma}_{\rm
  disp}$.  This implies that:

\be
\epsilon v^2_{\rm sn}  \times \Sigma_{\rm sfr} 
=\frac{1}{2} \Sigma_{\rm cold} \sigma_{\rm cold}^2 \kappa\, ,
\label{eq:epsilon}
\ee

\noindent
or, using the approximated value of $\kappa\simeq \sqrt{2}/\tau_{\rm orb}$:

\be
\Sigma_{\rm sfr} 
\simeq \frac{\Sigma_{\rm cold}}{ \tau_{\rm orb}} \times \frac{\sqrt{2}}{2}\left(\frac{\sigma_{\rm cold}}{\epsilon v^2_{\rm sn}}\right)^2
\, ,
\label{eq:epsilon2}
\ee

\noindent
This relation allows us to recast the interpretation of a unique
dynamical SK relation in terms of energy injection: it will hold as
long as, at fixed $\Sigma_{\rm sfr}$, 
the post-cooling efficiency of energy injection $\epsilon$
scales with the square of the gas vertical velocity dispersion,
i.e. the gas specific energy.  This can be seen in the other
direction: the specific energy of the gas $\sigma_{\rm cold}^2$ must
scale with the fraction of specific thermal energy that has time to
perform $PdV$ work before cooling radiates it away, $\epsilon v_{\rm
  sn}^2$.  This is a property that arises naturally from our
sub-resolution feedback model, at least for values of free
  parameters that do not widely differ from the fiducial ones selected
  in paper I; as in the case of pressure, we tested the validity of
this result with a large suite of {\muppi} simulations on the same
sets of initial conditions, and many combinations of parameters and
physical assumptions.  The result of an environment-dependent standard
SK and a unique dynamical SK holds in all cases where a disc
  efficiently heated by feedback and in a stationary state is
  obtained.

According to our interpretation, the fact that our four simulations
all lie on the same dynamical SK is {\em no} evidence that SFR is
directly determined by galaxy rotation, at variance with the
motivation that has been used to introduce this relation.  To make
this more clear, the dispersion rate of energy given in
equation~\ref{eq:diss} can be written, without loss of generality, as
$\dot{\Sigma}_{\rm disp}=3P\sigma_{\rm cold}$.  Pressure in our
simulations is well reproduced by equation~\ref{eq:Pfit}, that depends
on the kinematical state of the disc through $Q_{\rm tot}$, and then
on $\kappa$. As a result, the dynamical stationarity of the disc
induces a dependence of SFR on the epicyclic frequency, i.e. on the
orbital time. So, the fact that our four simulations all lie on the
same dynamical SK is telling us that they are stationary discs kept
out of vertical hydrostatic equilibrium by SN feedback, and not that
their SFR is directly determined by rotation or shearing - it is only
indirectly influenced by it through the dependence of pressure on
$Q_{\rm tot}$.  Indeed we noticed that some of our simulated discs
stay out of this dynamical SK, and this happens when, e.g., a bar
instability takes place. In this case the condition of stationarity is
violated until a new equilbrium configuration is reached. So these
discrepancies confirm our interpretation that simulations trace a
unique dynamical SK as long as they are in a stationary condition.

When run with the effective model of \cite{Springel03}, the three
galaxies (MW, DW, SH) show similar but not unique dynamical SK, the
differences being comparable to those in the standard SK
(section~\ref{section:vertical}).  This is a result of
inefficient thermal feedback in this model: as long as pressure is
not well fit by equation~\ref{eq:Pfit} and the injection of energy is marginal,
we do not expect galaxies to lie on a unique dynamical SK.

\section{Discussion}
\label{section:discussion}

One result of this paper is that our pressure-based model of
star formation produces a standard SK with a knee at a gas surface
density that depends on gas fraction in a way that resembles the
metallicity dependence proposed by \cite{Krumholz09b} and
\cite{Gnedin10}.  If the molecular fraction is regulated by pressure,
gas-rich galaxies become dominated by molecular gas at higher gas
surface densities.  It is easy to show that the gas surface density at
which $\Sigma_{\rm mol}$ equates $\Sigma_{HI}$, which we call
$\Sigma_{\rm eq}$, scales with the gas fraction $\mu=\Sigma_{\rm
  cold}/(\Sigma_{\rm cold} + \Sigma_\star)$, as:

\be
\Sigma_{\rm eq} = \frac{2P_0}{\pi G} \frac{\mu}{\mu + R (1-\mu)}\, .
\label{eq:Seq}
\ee

\noindent
Here $P_0$ is the normalization of equation~\ref{eq:fmol}.  In most
chemical evolution models, $\mu$ is related to metallicity; for
instance, in the simple closed-box model, $Z=Y\ln (1/\mu)$, where $Y$ is
the metal yield per stellar generation.  Then, a pressure-based
molecular fraction can mimic a metallicity dependence.  But the
maximum value of $\Sigma_{\rm eq}$ is reached for $\mu=1$, in which
case $\Sigma_{\rm eq, max} = \sqrt{2P_0/\pi G} \simeq 34\ {\rm
  M}_\odot\ {\rm pc}^{-2}$; this is the $\Sigma_{\rm eq}$ value of the
SH simulation, that has a 99 per cent gas fraction.  According to
\cite{Bolatto09}, $\Sigma_{\rm eq}$ for the Large Magellanic Cloud 
is similar to 34 {\surf},
while for 
the Small Magellanic Cloud it is $\sim$100 {\surf}.  This is
confirmed by a quick comparison with Krumholtz et al's model: an
increase in $\Sigma_{\rm eq}$ by a factor of 3, as large as the
difference between MW and SH, is obtained by a decrease of metallicity by
a similar factor, which is relatively modest.  The same conclusion can
be reached by considering the saturation $\Sigma_{\rm HI}$ value, that
reaches $\sim20$ {\surf} for our SH disc while, for instance,
\cite{Fumagalli10} report higher values for a few low metallicity
dwarf galaxies.  We conclude that a pressure-driven molecular fraction
cannot explain the whole observed range of variation of $\Sigma_{\rm
  eq}$.  Because the motivation for molecular fraction being directly
modulated by metallicity is very strong \citep[but see the comments
  by][]{MacLow10}, it is well conceivable to construct mixed scenarios
where equation~\ref{eq:fmol} is valid and the normalization $P_0$
depends on metallicity.

The search for a physical interpretation of our results lead us to an
expression for gas pressure in star-forming discs given by
equation~\ref{eq:Pfit}.  Discs subject to this pressure have some
interesting properties: for instance, neither $\Sigma_\star$ nor
$\sigma_\star$ appear in the expression of pressure (though the
gravity of stars enters in determining $\sigma_{\rm cold}$), gas
effective height is related to the ratio of gas velocity dispersion
and epicyclic frequency (equation~\ref{eq:Heff}) and the sound
crossing time of effective height is simply proportional to the
orbital time (equation~\ref{eq:tcross}).  Comparison with data is not
straightforward, as $\sigma_{\rm cold}$ includes a major contribution
from the sound speed of the hot phase (we give further comments below)
and direct pressure estimates are hard to obtain. Nonetheless, these
predictions can in principle be tested against observations of the
Milky Way and nearby galaxies, so they may constitute a basis for a
theory of the non-equilibrium vertical structure of discs effectively
heated by feedback.

Our simulated discs show total vertical velocity dispersions
(figure~\ref{fig:sigma}) that are well in excess of the $\sim10$
{\kms} value that is usually assumed to hold for discs.  However,
$\sigma_{\rm cold}$ in our simulations is dominated by the thermal
sound speed (equation~\ref{eq:scold}) computed at the particle
effective temperature, and this last quantity is determined by the
thermal content of the hot phase.  This means that $\sigma_{\rm cold}$
cannot be compared with the velocity dispersion of cold clouds in real
galaxies.  At the same time, vertical velocity dispersions of gas
particles (computed without the sound speed) of MW and DW simulations
show values that are in much better agreement with those measured in
THINGS galaxies by \cite{Tamburro09} (see figures 11 and 18 in paper
I).  In our simulations, multi-phase particles are composed by two
phases at the sub-grid level, but are seen as a single entity by the
SPH code.  This means that, as long as a multi-phase cycle goes on,
the hot phase is not free to leave the disc, while the cold phase is
pulled by the former.  This artificial entrainment results, at the
macroscopic level, in a velocity dispersion of gas particles that is
realistic and, from a detailed comparison of MW and MW\_HR, rather
stable with resolution in that part of the disc that is well resolved
in both simulations.  This means that our effective model is correctly
producing a gas disc that is thermally warm but kinematically colder.
It must be noticed that the hot coronae we produce around the discs
have temperatures $\sim3\times10^6$ K and densities $\sim10^{-2}$
cm$^{-3}$, so their presence is likely ruled out by X-ray observations
\citep[e.g.][]{Crain10}.  However, insertion of metal cooling would
change this energy balance in favour of kinetic energy, so we expect
this hot corona to be less prominent when chemical evolution is
properly taken into account.

Our results thus suggest that a complete modeling of a disc heated by
feedback must fully take into account the multi-phase nature of the
ISM, where the $\sim$kpc-height corona of warm/hot gas surrounding a
spiral galaxy may have an important dynamical role.  A step forward in
the modeling of a spiral disc subject to contiuous production and
dissipation of energy was recently taken by \cite{Elmegreen11}, who
addressed the stability of a disc where energy is continually injected
and dissipated over some multiple of the crossing time of turbulence.
However, in that paper only radial and tangential perturbations were
considered and vertical equilibrium was assumed.

The results presented in this paper rely on how well the vertical
structure of the disc is resolved in simulations.  In the MW and DW
cases effective disc heights are of the same order of the
gravitational softening (that was kept of the same order as the one
used in cosmological simulations with the same mass resolution), so
numerical convergence of the results should be demonstrated.  We
showed results for the MW\_HR simulation, and found no clear
dependence on resolution in any of our results.  Moreover, the light
disc forming out of the SH simulation has an effective height of
$\sim1$ kpc, with a gravitational softening of only 43 pc, so in this
case the vertical structure is well resolved.  We conclude that,
despite the vertical structure of the discs is barely resolved in some
of our simulations, the results should be not strongly affected by
resolution.

Due to the difficulty in performing measures of gas pressure,
observers typically use equation~\ref{eq:Pext} to estimate pressure
(often making strong assumptions on velocity dispersions), but its
validity has rarely been tested.  \cite{Koyama09} compared this
formula with simulations of turbulent ISM in a shearing disc.  In
their simulations the gas disc is assumed to be much thinner than the
stellar one, heating terms take account of cosmic rays, X-rays and
H$_2$ formation and destruction, while radiative feedback from massive
stars is modeled as localized increases of heating rate, but no SNe
are present.  Their spatial resolution is of order of $\sim$1 pc.
They found that midplane and mass-weighted pressures typically differ
by an order of magnitude, and that equation~\ref{eq:Pext} is very
close to the armonic mean of the two.  Interestingly, their midplane
pressure is a factor of three lower than $P_{\rm ext}$, which is what
we find for $Q_{\rm tot}=1$.  Our results are not comparable to their
simulation, due to the vastly different resolution and to the much
hotter state of our discs (they have $\sigma_{\rm cold} \sim5$
{\kms}).  They also proposed an improved analytic estimate of midplane
pressure; we compared it with our $P_{\rm sim}$ and found that it does
not improve much with respect to $P_{\rm ext}$.  Our simulations are
instead comparable to those of \cite{Tasker08} and \cite{Joung09},
that have resolutions of 25 and 2 pc respectively, include feedback
frm SNe and show
multi-phase structures of the ISM in broad agreement with the
assumptions made to design {\muppi}.  Unfortunately, they don't
explicitly address the question whether their pressure is well
reproduced by the external one of \cite{Elmegreen89}.

\section{Conclusions}
\label{section:conclusions}

In this paper we have shown how simulations based on the MUlti-Phase
Particle Integrator ({\muppi}) model
developed within the {\gtre} TreePM+SPH code, are able to give
predictions on the various SK relations discussed in the literature.
{\muppi} is based on the assumption, suggested by observations of
\cite{Blitz04,Blitz06}, that the molecular fraction is modulated by
pressure.  Moreover, it has the interesting feature of making thermal feedback
effective and able to heat discs.  So, the interest of this paper does
not rely only in the test of a specific sub-resoution model for star
formation but it allows
to understand what are the testable consequences of a pressure-based
molecular fraction in discs efficiently heated by SN feedback.

Our main conclusions are the following.

(i) A pressure-based molecular fraction produces an environmental-dependent standard SK
relation, owing to the fact that the relation
between pressure and gas surface density is modulated by gas fraction.
This variation is very similar to that found between spiral and dwarf
galaxies \citep{Bigiel08,Bigiel10}, and it could be interpreted as a metallicity
dependence, since gas fraction is typically related to metallicity in
most chemical evolution scenarios.  However, the variation in the
quantity $\Sigma_{\rm eq}$, the gas density at which the molecular
fraction is 1/2, cannot be larger than a factor of 3 between normal
spirals and gas-dominated discs, and values of $\Sigma_{\rm eq}>34$
{\surf} cannot be obtained in this framework.

(ii) We analyzed in detail the vertical structure of our simulated
galaxies, and found that hydrodynamical pressure is not well recovered by the
vertical hydrostatic equilibrium value of \cite{Elmegreen89}, because
kinematically hotter discs show higher pressure.  A better fit is
given by:

$$P_{\rm fit} = \frac{1}{6}\Sigma_{\rm cold} \sigma_{\rm cold} \kappa $$

\noindent
(equation~\ref{eq:Pfit}), that we interpret as the pressure of
a disc with continuous energy injection.  This expression allows to
connect the effective disc height with the gas velocity dispersion
(computed including kinetic and thermal energies) as $H_{\rm eff} = 3 \sigma_{\rm cold}/\kappa$
(equation~\ref{eq:Heff}).

(iii) Quite interestingly, our four simulated galaxies lie on the
  same (non-linear) dynamical SK relation, independently of their gas fraction.
  This is not a straightforward consequence of our assumptions, and
  was not expected.  It is worth recalling that a similar
  phenomenology was found by \cite{Daddi10b} in the very different
  context of $z\sim2$ star-forming galaxies.  We interpret this result
  as a manifestation of balance between energy injection from SNe and
  energy dissipation.  We have shown that, under the hypothesis that
  gas energy is dissipated both by cooling and by viscosity, and that
  the latter works on the timescale of one sound crossing time of the
  disc effective height, we can obtain for stationary discs a unique
  dynamical SK if the efficiency of energy injection after cooling
  scales with gas specific energy.  This is found to result from
  energy balance in our multi-phase particles under a wide range of
  cases.

(iv) Other results are of some interest.  The model follows well
the standard, molecular and HI SK relations, with a tight
molecular SK that is a straightforward consequence of the model
assumptions.  This molecular SK has a slope of $\sim1.4$,
marginally steeper than the 1.2 value found by \cite{Bigiel08} 
but in agreement with \cite{Liu11}.  The
scatter in the simulated relations is small, and this may hint that most
scatter is due to observational errors, if not to putting together
galaxies that lie on slightly different relations.  However, our
sub-resolution model gives by construction the average properties of
the ISM on scales that are similar to the 750 pc scale used to bin the
data, and this may be the reason for the low scatter.

These simulations show that energy injection by SNe is fundamental in
determining the structure of star-forming discs, and that the
  warm/hot phases created by stellar feedback may have an important
  role in disc dynamics.  Future observations will need to address
the issue of directly determining gas pressure in order to test
whether the usually assumed formula of \cite{Elmegreen89} or some
different estimates, like that provided by equation~\ref{eq:Pfit},
apply.

\section*{Acknowledgements}

We thank Frank Bigiel for providing his data on the SK relation of
spiral galaxies. Initial conditions for the simulations were kindly
provided by S. Callegari and L. Mayer (MW, MW\_HR, DW) and V. Springel
(SH).  Simulations were run at ``Centro Interuniversitario del
Nord-Est per il Calcolo Elettronico'' (CINECA, Bologna), with CPU time
assigned under an INAF/CINECA grant and under an agreement between
CINECA and University of Trieste, and at CASPUR, with CPU time
assigned with the ``Standard HPC grant 2009'' call.  We thank Anna
Curir, Bruce Elmegreen and Samuel Boissier for discussions.  We
acknowledge partial support by the European Commissions FP7 Marie
Curie Initial Training Network CosmoComp (PITN-GA-2009-238356) and by
grants ASI-COFIS, PRIN-MIUR 2007, PD51-INFN and PRIN-INAF 2009 titled
"Towards an italian network for computational cosmology".
K.D. acknowledges the support by the DFG Priority Programme 1177 and
additional support by the DFG Cluster of Excellence "Origin and
Structure of the Universe".

\bibliographystyle{mn2e}
\bibliography{master}

\label{lastpage}

\end{document}